\documentclass[]{spie} 

\usepackage{graphicx}
\usepackage{amsmath}
\usepackage{booktabs}
\usepackage{multirow}
\usepackage[colorlinks=true, allcolors=blue]{hyperref}

\title{MATISSE in the era of GRAVITY+.\\ on-sky performance of the mid-infrared
       spectro-interferometer with GPAO }

\author[a]{Florentin Millour}
\author[a]{Anthony Soulain}
\author[a]{St\'ephane Lagarde}
\author[a]{Alexis Matter}
\author[a]{Valentin Fleury}
\author[a]{Farin Drewes}
\author[a,b]{Mathis Houll\'e}
\author[a]{Jules Scigliuto}
\author[a]{Bruno Lopez}
\author[a]{Anthony Meilland}
\author[a]{Romain G. Petrov}
\author[a]{Philippe Berio}
\author[c,d]{Anthony Berdeu}
\author[e]{Guillaume Bourdarot}
\author[d]{Sylvestre Lacour}
\author[d]{Mathias Nowak}
\author[b]{Jean-Baptiste Le Bouquin}
\author[c]{Antoine M\'erand}
\author[e]{Frank Eisenhauer}
\author[c]{Julien Woillez}
\author[d]{Miguel Montarg\`es}
\author[e]{Diogo Ribeiro}
\author[e]{Taro Shimizu}
\author[c]{Jens Kammerer}
\author[f]{Xavier Haubois}
\author[e]{Stefan Gillessen}
\author[a]{Pierre Cruzal\`ebes}
\author[c]{Fr\'ed\'eric Gont\'e}
\author[g]{Walter Jaffe}
\author[a]{James Leftley}
\author[a]{Nicole Nesvadba}

\affil[a]{Laboratoire Lagrange, Universit\'e C\^ote d'Azur, Observatoire de la C\^ote d'Azur, CNRS,
          Bd. de l'Observatoire, CS 34229, 06304 Nice cedex 4, France}
\affil[b]{IPAG, Univ. Grenoble Alpes, CNRS, 38000 Grenoble, France}
\affil[c]{European Southern Observatory, Karl-Schwarzschild-Str.~2,
          85748 Garching bei M\"unchen, Germany}
\affil[d]{LIRA, Observatoire de Paris, Université PSL, CNRS, Sorbonne Université, Université de Paris Cité, 5 place Jules Janssen, 92195 Meudon, France}
\affil[e]{Max-Planck-Institut f\"ur extraterrestrische Physik, Giessenbachstrasse 1,
          85748 Garching, Germany}
\affil[f]{European Southern Observatory, Alonso de C\'ordova 3107, Vitacura,
          Casilla 19001, Santiago, Chile}
\affil[g]{Leiden Observatory, Leiden University, Niels Bohrweg 2,
          2333 CA Leiden, The Netherlands}

\authorinfo{Send correspondence to F.~Millour: \href{mailto:florentin.millour@oca.eu}%
{florentin.millour@oca.eu}}

\begin{document}
\pagestyle{plain}
\maketitle

\begin{abstract}
MATISSE, the mid-infrared spectro-interferometric imager of the Very Large
Telescope Interferometer (VLTI), has been feeding on the light corrected by the GRAVITY+ Adaptive Optics (GPAO) systems since the retirement of MACAO on the four Unit Telescopes in 2024. We report the assessment of the new MATISSE performance with GPAO, based on commissioning observations obtained in 2024 and 2025 and compared with archival MACAO observations of the same stellar sources. The instrumental visibility transfer function is more stable in the L band, while its typical value in the L and N bands is unchanged; the visibility signal-to-noise ratio is consistent with, or better than, the values reported at the initial MATISSE commissioning. The correlated flux increases by 20--30\% in GRA4MAT mode thanks to a better fringe tracking performance, and the total flux injected into the instrument increases by a factor of 1.3 to 1.4 (up to 1.7) in the L band at low spectral resolution, with no measurable loss in the N band. Combined with the new K-band fringe-tracking limits enabled by GPAO, this translates into a sensitivity gain of a factor 2 to 6 in the L and M bands, a confirmation of the N-band limits similar to the MACAO era. We then report the performances of MATISSE with the GPAO's laser guide stars in November and December 2025. The updated noise figures follow the exposure time calculator predictions in LM LOW and MED and come out better than them in N LOW, the L-band transfer function is stable at the few percent level between observing blocks an hour apart, and correlated fluxes, differential phases and closure phases are recovered down to a 10\,mJy target. Embedded young stellar objects and AGN are now within reach, thanks to the better performances on red targets brought by the lasers. We finally discuss the science cases opened by this gain and the case for a future upgrade of the instrument.
\end{abstract}


\section{INTRODUCTION}
\label{sec:intro}

MATISSE, the Multi AperTure mid-Infrared SpectroScopic Experiment, is the
second-generation mid-infrared instrument of the VLTI\cite{Lopez2022}. It
combines the beams of the four Unit Telescopes (UTs) or the four Auxiliary
Telescopes (ATs) and delivers spectrally dispersed fringes simultaneously in the L (3.0--3.9\,$\mu$m), M (4.5--5.0\,$\mu$m) and N (8.0--13\,$\mu$m) atmospheric windows, with spectral resolutions from $R\sim30$ (LOW) to $R\sim3300$ (HIGH+). Since 2019 observers using MATISSE have produced results on protoplanetary and circumstellar disks, evolved stars and the dusty structures of active galactic nuclei (Sect.~\ref{sec:science}).

MATISSE has no fringe tracker of its own. The MATISSE consortium set out its requirements for an external fringe tracker: four telescopes, chopping compatibility, off-axis capability, a tracking sensitivity of $K>10$ with a goal of $K>12$, a coherencing sensitivity of $K>12$ with a goal of $K>14$, and a tracking accuracy of 180\,nm RMS over one minute. The driver was that the L-band MED and HIGH modes are unusable without external stabilisation: with the UTs the SNR${}=3$ limit per frame was estimated at 2 and 4\,Jy for a 50\,ms exposure, against 0.3 and 0.4\,Jy with a fringe tracker and a 250\,ms exposure. The GRA4MAT mode, in which the GRAVITY K-band fringe tracker stabilises the optical path difference for MATISSE, was commissioned over 2019--2023 and answered these requirements. It improved the L- and M-band sensitivity by an order of magnitude and made all the spectral modes of the instrument usable\cite{Woillez2024}. 

Besides the fringe tracking, the second lever on MATISSE performance is
the wavefront correction delivered to the VLTI by its adaptive optics. The GRAVITY+ project has replaced the MACAO adaptive optics systems of the four UTs by GPAO, a set of four high-order adaptive optics systems offering both visible and infrared natural guide star (NGS) modes and, since 2025, laser guide star (LGS) operation\cite{GravityPlus2026, GravityPlus2022, Millour2024}. GPAO was designed and built by a French-German consortium: the system and the deformable mirrors in Grenoble, the wavefront sensors at MPE, the real-time computer in Paris, and the integration in Nice. The laser guide stars are provided by ESO.

GPAO feeds all VLTI instruments, so its benefits are not restricted to GRAVITY. In this paper we quantify what MATISSE gains from it. Section~\ref{sec:vis} describes the MACAO/GPAO comparison dataset and presents the measured transfer function, signal-to-noise ratio, correlated flux and injected flux. Section~\ref{sec:limits} gives the resulting sensitivity limits in NGS mode. Section~\ref{sec:lgs} reports the recommissioning of MATISSE with the laser guide stars, and Section~\ref{sec:science} places these numbers in the context of recent and forthcoming MATISSE science.

\subsection*{What to expect from GPAO}

Four ingredients are expected to propagate the better GPAO wavefront correction into the MATISSE data: a higher and more stable Strehl ratio delivered to the VLTI beams; a more stable fringe tracking with GRAVITY, hence a lower rejection rate of MATISSE frames from the GRA4MAT telemetry; more flux injected into the MATISSE spatial filters; and finally, with the laser guide stars, a better beam control, and therefore a better accuracy, on faint sources for which the NGS mode of GPAO would not be sensitive-enough.

On the MATISSE side, the corresponding observable expected improvements are a better stability of the instrumental visibility (in standalone and GRA4MAT modes, mostly in the L and M bands), a better K-band fringe tracking sensitivity of  GRA4MAT, better overall LM- and N-band sensitivities, and access to redder and more embedded targets. The following sections quantify each of them.

\section{MACAO vs. GPAO (NGS)}
\label{sec:vis}

\subsection{The GPAO NGS campaign}
\label{sec:obs}

We gathered the MATISSE NGS data over one continuous campaign, from the GPAO
commissioning nights of 14 and 17--18 November and 11, 12 and 14 December 2024
to the technical runs of 9--10 February, 14 and 15 March, 8 June and 3 August
2025, all on the UTs\cite{GpaoNgsReport}. The commissioning nights were used
in standalone and in GRA4MAT, and the technical runs covered many spectral
resolutions of MATISSE in both modes. We use the first part of the campaign for
the MACAO/GPAO comparison presented below, and the whole of it for the
performance update of Sect.~\ref{sec:perfupdate}.

We compare the MATISSE observations obtained with GPAO to archival MATISSE
observations of the same calibrator or science target obtained with MACAO, in
the same spectral mode and, whenever possible, with the same telescope
configuration. The targets used here span two decades in mid-infrared
brightness:

\begin{itemize}
    \item HD~13763, a bright calibrator at 16.9\,Jy in L and 2.7\,Jy in N according to
the MDFC\cite{Cruzalebes2019}, is used for the transfer-function and
injected-flux comparison; 
\item UCAC2~3446325, at 0.2\,Jy in L, probes the low-flux
regime;
\item $\beta$~Pictoris, at 11.6\,Jy in L, for the MED-resolution
correlated flux;
\item HD~36255 for the high-resolution (VHIGH/HIGH+) flux
comparison.
\end{itemize}

We compute the signal-to-noise ratio (SNR) with the same method as in the original
MATISSE commissioning work\cite{Lopez2022, Petrov2018, Petrov2020}, so that the
numbers presented below can be compared directly with the pre-GPAO performance.
Over these runs, we obtained usable data down to 200\,mJy in LOW resolution in
standalone mode, 8\,mJy in MED with GRA4MAT in the narrow off-axis mode, and
6\,Jy in HIGH+ with GRA4MAT\cite{GpaoNgsReport}.

\subsection{Instrumental visibility and SNR in the L band: the same average value as before, but more stable}

\begin{figure}[htbp]
  \centering
  \includegraphics[width=0.48\textwidth]{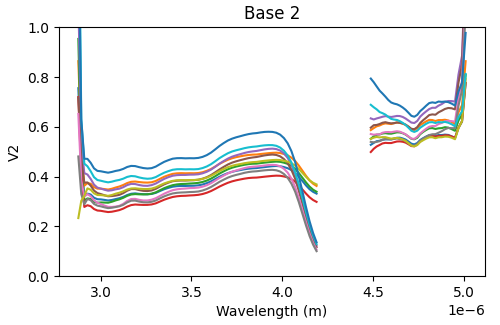}\hfill
  \includegraphics[width=0.48\textwidth]{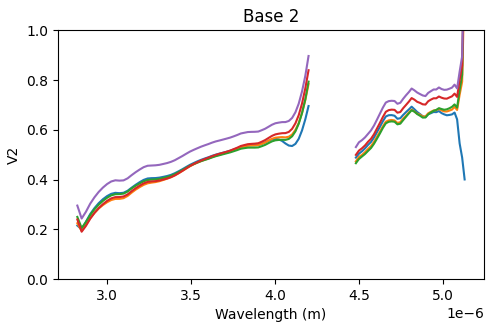}
  \caption{Instrumental visibility of MATISSE in the L band, LOW resolution, on
  HD~13763 (16\,Jy). \emph{Left:} MACAO, 2019. \emph{Right:} GPAO, 2024. The
  typical value is unchanged; the stability over time is improved with GPAO.}
  \label{fig:visL}
\end{figure}

Figure~\ref{fig:visL} compares the instrumental (raw) visibility measured on
HD~13763 in LOW resolution, in 2019 with MACAO and in 2024 with GPAO. The
typical level of the transfer function is unchanged, while its dispersion over
the duration of the observing sequence is reduced with GPAO. This gain in
stability results from the adaptive optics and from the improved fringe-tracking
performance that follows from it. It matters for the calibration accuracy of
MATISSE, since the transfer function is interpolated between calibrator
observations.

The same comparison at lower flux (UCAC2~3446325, 0.2\,Jy in L) shows that the
overall behaviour of MATISSE is preserved: the instrument responds in the same
way using the new adaptive optics.

\begin{figure}[htbp]
  \centering
  \includegraphics[width=0.85\textwidth]{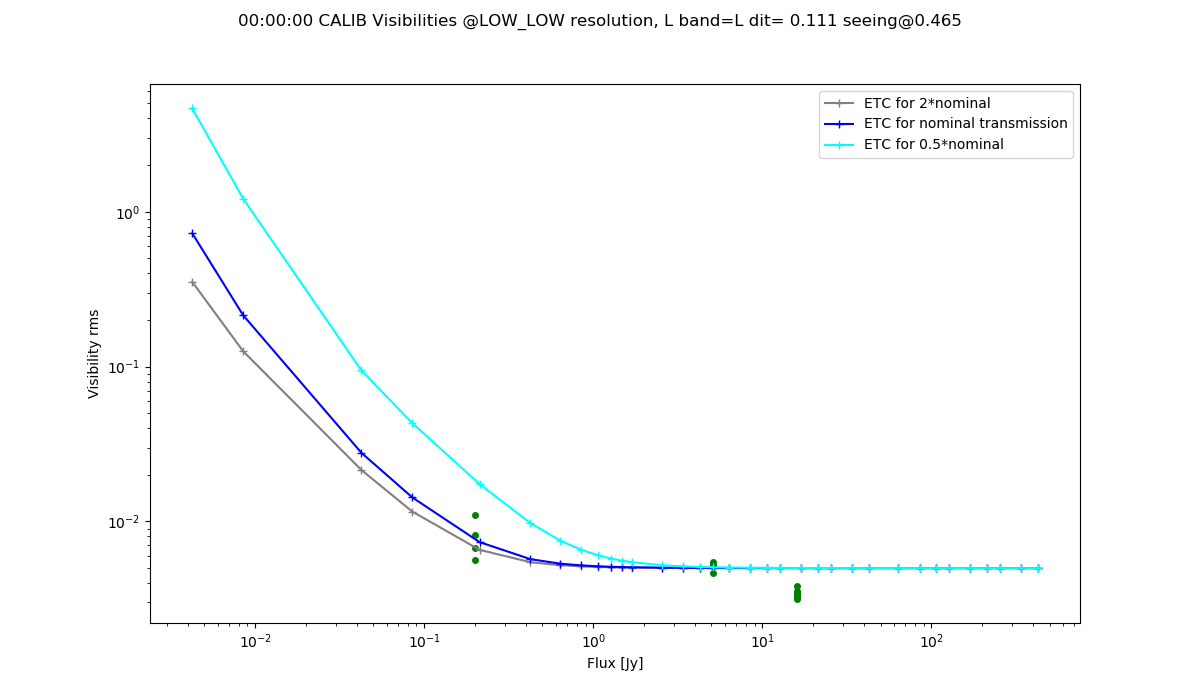}
  \caption{Visibility signal-to-noise ratio assessment in the L band. The
  MATISSE SNR with GPAO is consistent with the performances measured at
  commissioning.}
  \label{fig:snrL}
\end{figure}

We estimated the visibility SNR (Fig.~\ref{fig:snrL}) with the method of
Ref.~\citenum{Petrov2020}, and find it consistent with the previously measured
performances.

\subsection{Instrumental visibility and SNR in the N band: the same as before}

\begin{figure}[htbp]
  \centering
  \includegraphics[width=0.4\textwidth]{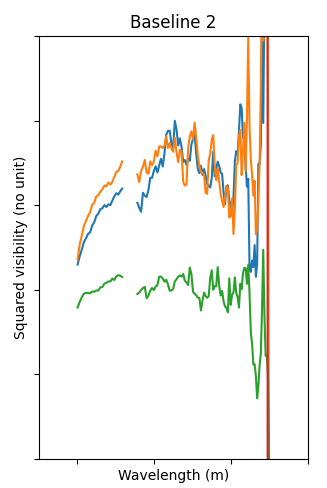}\hfill
  \includegraphics[width=0.4\textwidth]{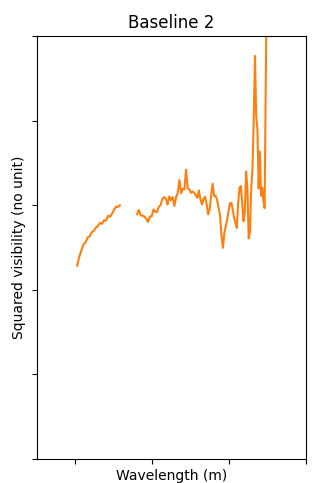}
  \caption{Instrumental visibility in the N band on HD~13763 (2\,Jy).
  \emph{Left:} MACAO. \emph{Right:} GPAO. The typical value is the same as with
  MACAO.}
  \label{fig:visN}
\end{figure}

\begin{figure}[htbp]
  \centering
  \includegraphics[width=0.85\textwidth]{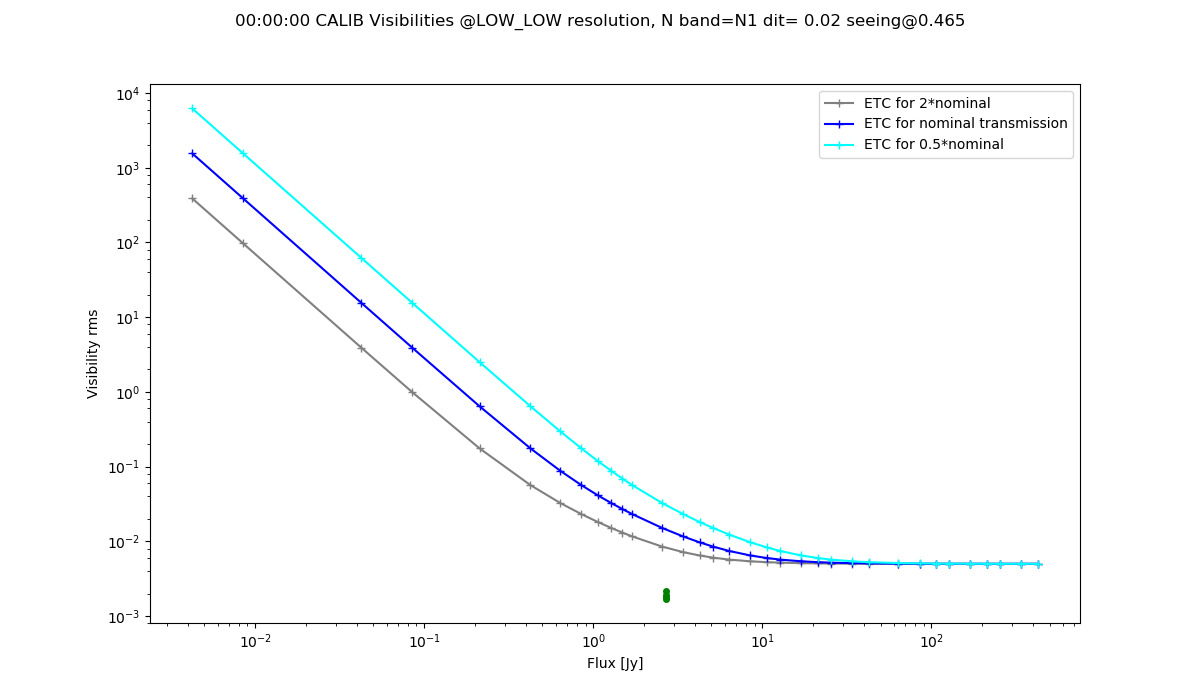}
  \caption{Visibility signal-to-noise ratio assessment in the N band. The
  MATISSE SNR is consistent with the previous performance, or better. This plot
  relies on a single chopping measurement (see text).}
  \label{fig:snrN}
\end{figure}

The N band is much less sensitive to the quality of the wavefront correction: at
10\,$\mu$m the UTs are close to the diffraction limit even with a modest Strehl
ratio in the visible. Figure~\ref{fig:visN} confirms this expectation: on HD~13763 the N-band
visibility is close to 0.6 with both the MACAO and GPAO datasets. A second
target observed with GPAO, HD~36558 (5.1\,Jy), gives 0.4, so the transfer
function varies by about $\pm10$\% between the two.
Figure~\ref{fig:snrN} shows an N-band visibility SNR consistent with, or slightly
better than, the previous performances. This rests on a single chopping measurement, the N-band configuration having
been available only on the last night of the run.

\subsection{Correlated flux}
\label{sec:corrflux}

\begin{figure}[htbp]
  \centering
  \includegraphics[width=0.4\textwidth]{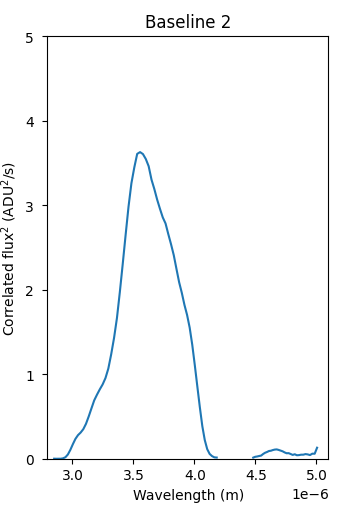}\hfill
  \includegraphics[width=0.4\textwidth]{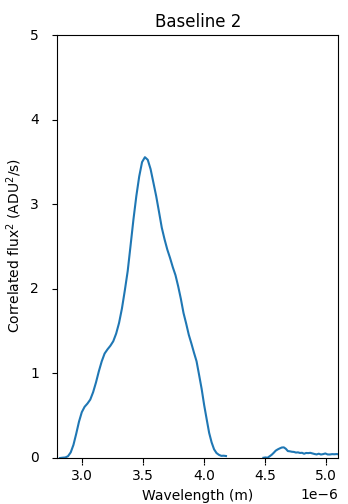}
  \caption{Correlated flux in the L band, LOW resolution, on UCAC2~3446325
  (0.2\,Jy). \emph{Left:} MACAO. \emph{Right:} GPAO. Standalone values are
  unchanged; GRA4MAT gains 20--30\%.}
  \label{fig:corrfluxfaint}
\end{figure}

\begin{figure}[htbp]
  \centering
  \hfill\includegraphics[width=0.935\textwidth]{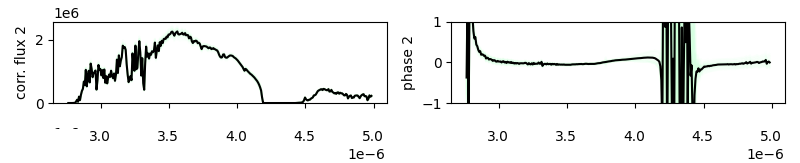}\\
  \hfill\includegraphics[width=0.99\textwidth]{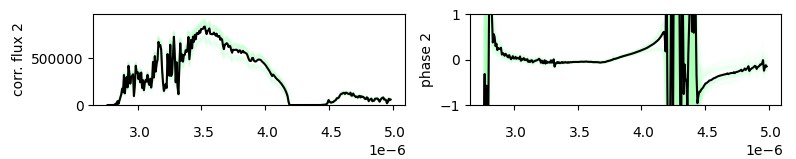}
  \caption{Correlated flux in the L band, MED resolution, on $\beta$~Pic
  (11.6\,Jy). \emph{Top:} MACAO. \emph{Bottom:} GPAO. The correlated flux of
  the star is larger with GPAO by 13 to 30\% depending on the baseline.}
  \label{fig:corrfluxmed}
\end{figure}

The correlated flux benefits directly from the improved fringe tracking. At low
flux in the L band (UCAC2~3446325, 0.2\,Jy, Fig.~\ref{fig:corrfluxfaint}), the
standalone mode delivers the same typical correlated flux as with MACAO, while
the GRA4MAT mode gains 20 to 30\%. At medium resolution on a bright target
($\beta$~Pic, 11.6\,Jy, Fig.~\ref{fig:corrfluxmed}), the correlated flux of the
star, renormalised to a common integration time, is larger with GPAO by 13 to
30\% depending on the baseline, with a mean of 22\%, while the correlated flux
of the planet is unchanged\cite{GpaoNgsReport}.

\subsection{Injected (total) flux}
\label{sec:flux}

\begin{figure}[htbp]
  \centering
  \includegraphics[width=0.48\textwidth]{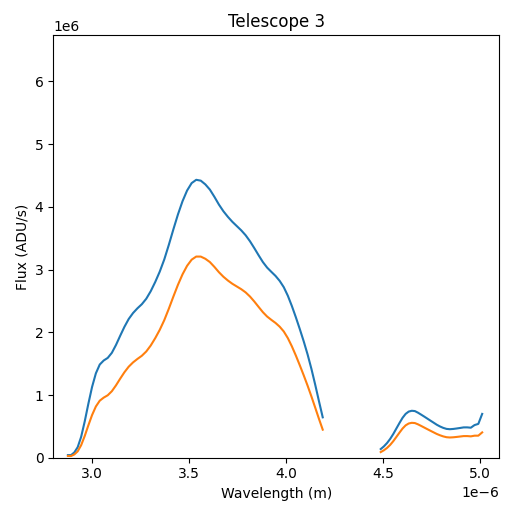}\hfill
  \includegraphics[width=0.48\textwidth]{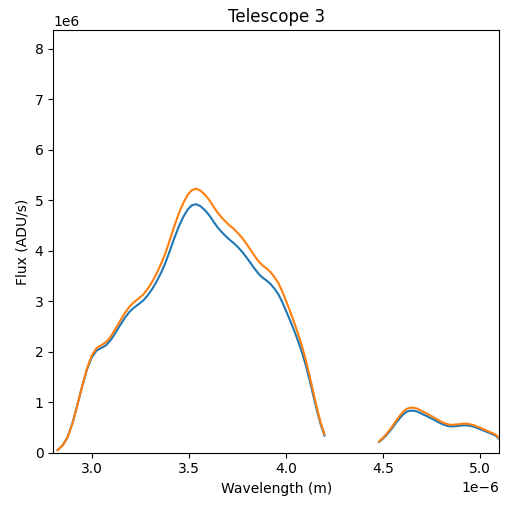}
  \caption{Total (injected) flux in the L band, LOW resolution, on HD~13763.
  \emph{Left:} MACAO. \emph{Right:} GPAO. Note the different flux scales on the vertical axes. Average flux increase of 1.3 to 1.4
  per telescope, up to 1.7 on the best exposure.}
  \label{fig:totfluxL}
\end{figure}

The total flux recorded by MATISSE is a direct proxy for the injection
efficiency, and therefore for the Strehl ratio delivered by the adaptive optics.
On HD~13763 in LOW resolution (Fig.~\ref{fig:totfluxL}), the flux averaged over
3.3--4.0\,$\mu$m increases by a factor of 1.3 to 1.4 depending on the
telescope, with a maximum single-exposure gain of 1.7
(Table~\ref{tab:fluxgain}).
\begin{table}[htbp]
  \centering
  \caption{Average flux over 3.3--4.0\,$\mu$m on HD~13763, per telescope, in
  detector units, averaged over the two available exposures for each adaptive
  optics system\cite{GpaoNgsReport}. The maximum gain is computed on the single
  best exposure of each pair.}
  \label{tab:fluxgain}
  \begin{tabular}{lcccc}
    \toprule
     & Telescope 1 & Telescope 2 & Telescope 3 & Telescope 4 \\
    \midrule
    MACAO        & $4.3\times10^{7}$ & $3.9\times10^{7}$ & $3.1\times10^{7}$ & $4.8\times10^{7}$ \\
    GPAO         & $6.2\times10^{7}$ & $5.4\times10^{7}$ & $4.2\times10^{7}$ & $6.1\times10^{7}$ \\
    \midrule
    Average gain & 1.4 & 1.4 & 1.3 & 1.3 \\
    Maximum gain & 1.7 & 1.5 & 1.6 & 1.3 \\
    \bottomrule
  \end{tabular}
\end{table}

\begin{figure}[htbp]
  \centering
  \includegraphics[width=0.48\textwidth]{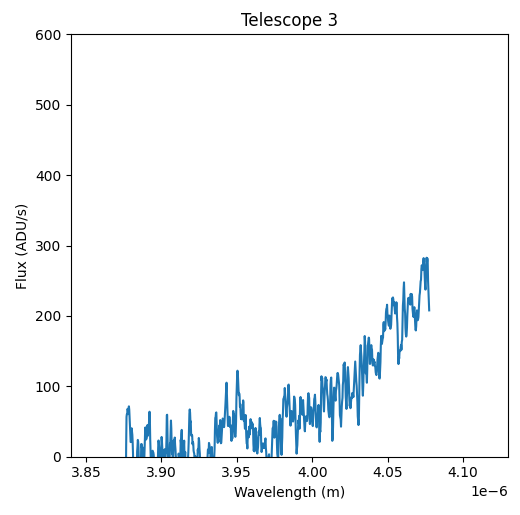}\hfill
  \includegraphics[width=0.48\textwidth]{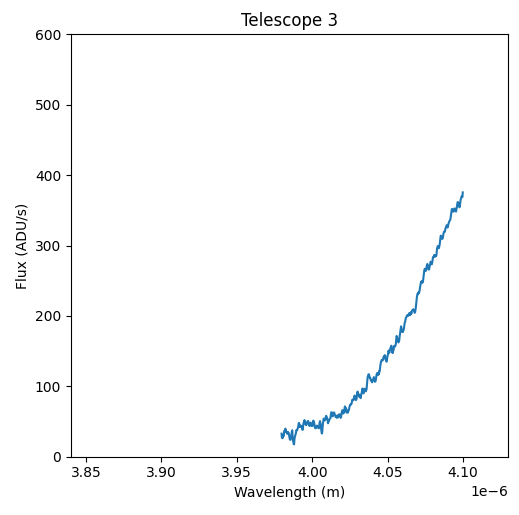}
  \caption{L-band spectra of HD~36255 at high spectral resolution, per
  telescope. \emph{Left:} MACAO, DIT $=0.6$\,s. \emph{Right:} GPAO, DIT
  $=10$\,s. The flux level is unchanged, while the longer exposure permitted by
  the stabilised fringes yields a much cleaner spectrum.}
  \label{fig:totfluxLHIGH}
\end{figure}

At medium resolution the increase is also present, although the measurement
saturates around 10\,Jy. At very high resolution (VHIGH/HIGH+) no flux increase is measured: on HD~36255
(6.1\,Jy), averaged over 4.00--4.07\,$\mu$m, the per-telescope gains are 1.04,
0.86, 0.78 and 1.01\cite{GpaoNgsReport}. The SNR
still improves, because the better fringe tracking allows much longer detector
integration times. Figure~\ref{fig:totfluxLHIGH} illustrates this on HD~36255,
observed with a detector integration time (DIT) of 0.6\,s with MACAO and 10\,s with GPAO.

In the N band, no flux increase is expected, and the measurement on HD~13763
shows neither a gain nor a loss. The GPAO upgrade leaves the N-band arm of
MATISSE unaffected.

\section{UPDATED SENSITIVITY LIMITS WITH GPAO NGS}
\label{sec:limits}

\subsection{Natural guide star mode}

Table~\ref{tab:ngs} gives the sensitivity limits offered in Period~117 for
GRA4MAT with the UTs, with GPAO in visible NGS mode. In the L and M bands the
gain with respect to the pre-GPAO limits ranges from a factor 2 at LOW
resolution to a factor 6--7 at MED and HIGH resolution. In the N band the
existing limits are confirmed rather than improved, in agreement with the
measurements of Sect.~\ref{sec:flux}.

The K-band limits of the fringe tracker as used by GRA4MAT were revised on the
basis of the February 2025 technical run, during which fringes were tracked on
calibrators from $K=10.5$ to $K=11.5$ and on two AGN at $K\simeq10$. The
proposed limits are $K=10$ in regular conditions (70\% best), $K=11$ in good
conditions (30\% best) and $K=11.5$ in the very best (10\%), against $K=9$,
$K=10$ and $K=10.5$ before\cite{GpaoNgsReport}, a gain of one magnitude. These
should not be confused with the limits of the GRAVITY fringe tracker as used by
GRAVITY itself, which are about one and a half magnitudes fainter.

The same run led to the L-band correlated flux and closure phase limit being
lowered to 0.01\,Jy, from 20\,mJy previously. Reaching it requires chopped
frames: on the $K=11$ calibrator EC~11107-3237 ($L=11$\,mJy) the non-chopped
data show visibility and flux biases that make them unusable, whereas the
chopped frames are consistent across exposures\cite{GpaoNgsReport}.

\begin{table}[htbp]
  \centering
  \caption{MATISSE Period~117 sensitivity limits, GRA4MAT with the UTs, GPAO
  NGS-VIS. Fluxes in Jy. The gain factors with respect to the pre-GPAO limits
  are given in parentheses for the correlated flux and differential phase
  columns.}
  \label{tab:ngs}
  \resizebox{\textwidth}{!}{%
\begin{tabular}{lccccccc}
    \toprule
    \multirow{2}{*}{Spectral mode} & \multicolumn{2}{c}{Lowest L flux (Jy)}
      & \multicolumn{2}{c}{Lowest M flux (Jy)}
      & \multicolumn{2}{c}{Lowest N flux (Jy)} \\
    \cmidrule(lr){2-3}\cmidrule(lr){4-5}\cmidrule(lr){6-7}
      & Corr. flux \& diff. phase & Closure phase
      & Corr. flux \& diff. phase & Closure phase
      & Corr. flux \& diff. phase & Closure phase \\
    \midrule
    LOW ($R\sim35$)    & 0.01 ($\times$2) & 0.03 & 0.03 ($\times$2) & 0.04
                       & 0.1 (confirmed)  & 0.1 \\
    MED ($R\sim500$)   & 0.04 ($\times$6) & 0.06 & 0.35 ($\times$3) & 0.45
                       & not offered      & not offered \\
    HIGH ($R\sim1000$) & 0.12 ($\times$7) & 0.16 & not offered      & not offered
                       & 0.5 ($R\sim220$) & 0.7 ($R\sim220$) \\
    HIGH+ ($R\sim3300$)& 6                & 8    & 6                & 8
                       & not offered      & not offered \\
    \bottomrule
  \end{tabular}}
\end{table}

\subsection{Updated performance estimates from the 2025 runs}
\label{sec:perfupdate}

The limits of Table~\ref{tab:ngs} are the values offered to the community. They
derive from a dedicated analysis of the February to August 2025 runs, whose
outcome is reproduced in Table~\ref{tab:perfupdate}. It also produced a
separate figure for the visibility, which the call for proposals does not quote
alongside the correlated flux, and a breakdown of the N band into its three
low-resolution sub-bands.

The limits are defined as the minimum coherent flux giving, per spectral
channel and per one-minute exposure, an accuracy of 0.1 on the visibility,
4$^\circ$ on the differential phase and 5$^\circ$ on the closure phase. The
4$^\circ$ criterion corresponds to a signal-to-noise ratio of 10 on the
coherent flux. The visibility limit includes the broadband photometric and
calibration errors and the closure phase limit its own broadband calibration
errors, both taken from Tables~5 and 6 of Ref.~\citenum{Lopez2022} and combined
following its Eq.~A.9.

\begin{table}[htbp]
  \centering
  \caption{Updated MATISSE performance estimates from the February to August
  2025 GPAO runs. Fluxes are the minimum coherent flux per spectral channel and
  per one-minute exposure meeting the criteria given in the text.}
  \label{tab:perfupdate}
  \resizebox{\textwidth}{!}{%
\begin{tabular}{llccccl}
    \toprule
    Mode & Band, resolution & Visibility & Corr. flux \& diff. phase
         & Closure phase & DIT & Faintest target \\
    \midrule
    GRA4MAT & L-HIGH  & 70\,mJy  & 120\,mJy & 160\,mJy & 10\,s    & 700\,mJy (extrapolated) \\
    GRA4MAT & L-MED  & 50\,mJy  & 35\,mJy  & 60\,mJy  & 10\,s    & 100\,mJy \\
    GRA4MAT & M-MED  & 180\,mJy & 350\,mJy & 450\,mJy & 10\,s    & 30\,mJy \\
    GRA4MAT & M-LOW  & 110\,mJy & 30\,mJy  & 35\,mJy  & 1.3\,s   & 10\,mJy \\
    standalone & M-LOW & 110\,mJy & 40\,mJy & 40\,mJy & 0.111\,s & 100\,mJy (300 chopped) \\
    \midrule
    GRA4MAT & N1-LOW & --       & 100\,mJy & 400\,mJy & 0.02\,s  & near the limit, no bias \\
    GRA4MAT & N2-LOW & --       & 350\,mJy & 500\,mJy & 0.02\,s  & idem \\
    GRA4MAT & N3-LOW & --       & 550\,mJy & 800\,mJy & 0.02\,s  & idem \\
    \bottomrule
  \end{tabular}}
\end{table}

In the M band the standalone and GRA4MAT limits coincide, because there is no
gain from the longer integration time in a regime dominated by background
photon noise, and because the piston is measured in L, which removes the
low-coherent-flux bias that would otherwise affect the standalone mode. In the
N band the GRA4MAT measurements confirm that the mode observes without that
bias.

Three caveats apply. Exposures are occasionally lost to a partial failure of
the adaptive optics or of the fringe tracking, so the data have to be
inspected. Phase wrapping is still present in the pipeline output and has to be
corrected in post-processing, although a version of the pipeline in test
addresses it. The N band retains residual pistons, most likely from an
imperfect correction of the chromatic piston difference between the K and N
bands, which can be removed at the analysis stage.

\section{MATISSE IN THE LASER GUIDE STAR ERA}
\label{sec:lgs}

The measurements reported so far were obtained with GPAO in natural guide star
mode. The laser guide stars were commissioned at the end of 2025, and MATISSE
was recommissioned alongside them\cite{MatisseLgsReport}.

\subsection{Goals of the recommissioning}

The consortium set three goals for this campaign: recommission MATISSE with the
LGS mode in view of the improved sensitivity of GPAO+LGS; propose limiting
magnitudes for MATISSE with LGS to ESO; demonstrate on faint representative
targets that usable MATISSE data can be obtained.

The observing plan covers a sequence of calibrators from $R=12$ down to
$R=18$--19 to bracket the limit, young stellar objects and AGN that could not
be observed until now in a specific mode, the limiting sensitivity in the LOW-LM
and MED-LM modes with LOW-N and HIGH-N in parallel, the limiting magnitude of
GRA4MAT set by the number of fringe jumps, and the transfer function stability
on brighter targets. Two questions are left open for later runs: the standalone
mode, for targets too faint in K for GRA4MAT but bright enough in L, M or N,
and isoplanatism, for which repeating the sensitivity measurements at several
guide star separations on the same stars as GRAVITY would give the
MATISSE/GRAVITY scaling law.

\subsection{Observations}

Table~\ref{tab:lgslog} lists the sequence. We got the first light in LOW on the
night of 4--5 November 2025. MATISSE was then used again on the nights of
7--8 November and 8--9 November, and in two slots of the December run, the
second of them in poor conditions with an average
$\tau_0 < 3$\,ms. All observations were carried out on the UTs in GRA4MAT mode.
Over the two December nights, which were run in an operations-driven rather
than commissioning-driven way, the observing cadence was about one target every
half hour.

\begin{table}[htbp]
  \centering
  \caption{MATISSE sequence during the GPAO+LGS commissioning runs. $K$ is the
  2MASS magnitude and $F_L$, $F_N$ the median L- and N-band fluxes from the
  MDFC\cite{Cruzalebes2019}, both retrieved through the CDS; where the target is
  absent from the MDFC the value of the commissioning report is given instead.
  $G_{\rm RP}^{\rm TT}$ is the magnitude of the tip-tilt star, which is not the
  science target (see text). The DIMM column is the median of the Paranal ASM
  DIMM-2016 seeing (500\,nm, zenith) over the time each target was observed by
  MATISSE; it is quoted only for the intervals the night logs identify as
  MATISSE observations, on nights whose date is unambiguous.}
  \label{tab:lgslog}
  \resizebox{\textwidth}{!}{%
\begin{tabular}{llccccccl}
    \toprule
    Night & Target & $K$ & $F_L$ (Jy) & $F_N$ (Jy) & $G_{\rm RP}^{\rm TT}$
          & DIMM ($''$) & Mode & DIT \\
    \midrule
    7--8 Nov  & HD~30900    & 7.20  & 0.51      & 0.052     & 8.7  & 0.50 & acquisition & -- \\
              & CD-22\,624  & 11.02 & 0.019     & 0.0022    & 10.0 & 0.69 & LOW-LM/LOW-N & 1.3\,s \\
              & CD-27\,1225 & 11.09 & 0.012     & 0.0014    & 15.6 & 0.62 & LOW-LM/LOW-N & 1.3\,s \\
              & NGC~1068    & --    & 1$^{a}$   & 10$^{a}$  & --   & 0.75 & MED-LM/HIGH-N & 1.3\,s \\
              & CD-29\,1424 & 11.09 & 0.012     & 0.0013    & 16.5 & 0.95 & LOW-LM/LOW-N & 1.3\,s \\
              & HD~294236   & 11.33 & 0.0094    & 0.0012    & 14.9 & 1.00 & LOW-LM/LOW-N & 1.3\,s \\
    \midrule
    8--9 Nov  & HD~285851   & 7.25  & 0.39      & 0.043     & 9.7  & 0.50 & MED-LM/LOW-N & 1.3\,s \\
              & Haro~6-13   & 8.10  & 0.6$^{a}$ & 0.5$^{a}$ & 15.0 & 0.56 & MED-LM/LOW-N & 1.3\,s \\
    \midrule
    7--8 Dec  & HL~Tau      & --    & 2.5$^{a}$ & 9$^{a}$   & --   & --   & MED-LM/LOW-N & 5\,s \\
              & HL~Tau      & --    & 2.5$^{a}$ & 9$^{a}$   & --   & --   & VHIGH-L/LOW-N & 10\,s \\
              & HD~26546    & --    & 10$^{a}$  & 1.6$^{a}$ & --   & --   & MED-LM/LOW-N & 5\,s \\
    \midrule
    11--12 Dec  & HD~7402     & 5.26  & 2.37      & 0.34      & 6.8  & 0.63 & MED-LM/LOW-N, NGS & 5\,s \\
              & HD~1103     & --    & --        & --        & 16.9 & --   & aborted & -- \\
              & HD~7402     & 5.26  & 2.37      & 0.34      & 6.8  & --   & MED-LM/LOW-N, LGS & 5\,s \\
              & HD~29048    & --    & --        & --        & 16.4 & --   & aborted & -- \\
              & HD~40466    & --    & 0.8$^{a}$ & 0.1$^{a}$ & 7.9  & --   & MED-LM/LOW-N & 5\,s \\
              & HD~290139   & --    & 0.1$^{a}$ & 0.07$^{a}$& 9.2  & --   & MED-LM/LOW-N & 5\,s \\
              & HD~16589    & --    & 2.6$^{a}$ & 0.36$^{a}$& 5.9  & --   & MED-LM/LOW-N & 5\,s \\
              & HD~29054    & --    & 1.4$^{a}$ & 0.21$^{a}$& 4.2  & --   & MED-LM/LOW-N & 5\,s \\
    \bottomrule
  \end{tabular}}

  \vspace{1mm}
  {\footnotesize $^{a}$ value from the commissioning report.}
\end{table}

We retrieved the fluxes and magnitudes of the November targets in Table~\ref{tab:lgslog} from the CDS rather than taking them from the raw night logs, for two reasons. First, the $K$ magnitudes entered in the logs as fringe-tracking magnitudes agree with the 2MASS values of the science targets to better than 0.1\,mag in every case, which confirms that MATISSE was fringe-tracking on the target itself. Second, the L fluxes recorded in the raw logs for the two faintest calibrators are an order of magnitude too high: 0.1\,Jy for both CD-27\,1225 and HD~294236, where the MDFC gives 12 and 9.4\,mJy. The MDFC values agree with the commissioning report, which quotes 0.01\,Jy for both.

The tip-tilt magnitudes need care. For CD-27\,1225, CD-29\,1424 and HD~294236, Gaia~DR3 gives $G_{\rm RP} = 11.45$, 11.43 and 11.59 for the targets themselves, well brighter than the values in Table~\ref{tab:lgslog}. The low-order sensing was therefore performed on separate field stars, as expected in LGS mode; the commissioning report gives their separations as 27.7$''$ for CD-27\,1225 and 21.6$''$ for HD~294236. These values cannot be checked against
a catalogue without the corresponding observing blocks.

Three presets did not produce usable data, and all three failed in the
low-order loop. On CD-22\,624 the loop locked on the science star itself. On
CD-29\,1424 ($G_{\rm RP}^{\rm TT} = 16.5$) fringes were found but could not be
tracked. On the night of 11--12 December, the preset on HD~1103 was aborted
after a wavefront sensor freeze, the log noting that the tip-tilt star at
$G_{\rm RP} = 16.9$ was too faint, and the preset on HD~29048
($G_{\rm RP}^{\rm TT} = 16.4$) was abandoned when UT3 could not find the laser
guide star. Every tip-tilt reference fainter than $G_{\rm RP} \simeq 16$ thus
failed, while every successful one was brighter than 15.6. The seeing was also
the worst of its night for CD-29\,1424, at 0.95$''$, so guide-star magnitude
and atmospheric conditions are partly degenerate in this dataset; separating
the two effects would require a sequence at fixed seeing.

\subsection{Sensitivity against the exposure time calculator}
\label{sec:lgssens}

We compared our on-sky measurements in the L and N bands to the official
MATISSE exposure time calculator (ETC). We extracted the fundamental noise
level, i.e. the RMS in the spectral direction of the visibility, differential
phase and closure phase, with the method of Ref.~\citenum{Petrov2020}, and
compared it to the ETC predictions for three values of the global transmission including the effect of the Strehl ratio: the
theoretical transmission of the performance analysis report, half of it, and
twice it. For the visibility and the closure phase, the transmission case that
best fits the measurements is then used to generate two further curves that
include the broadband calibration and photometric errors, one for good and one
for poor seeing. In GRA4MAT mode the good-seeing curve is the relevant one:
once the GRAVITY fringe tracker is locked, the dependence of the MATISSE
instrumental visibility on seeing nearly disappears.

\begin{figure}[htbp]
  \centering
  \includegraphics[width=0.5\textwidth]{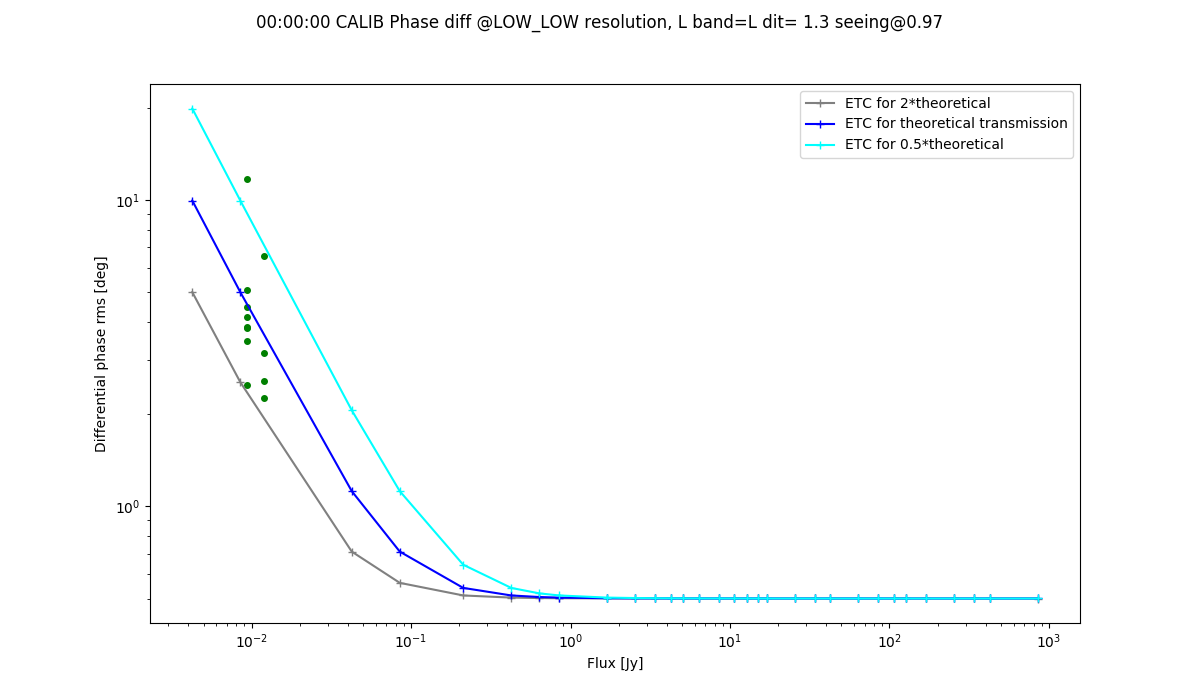}\hfill
  \includegraphics[width=0.5\textwidth]{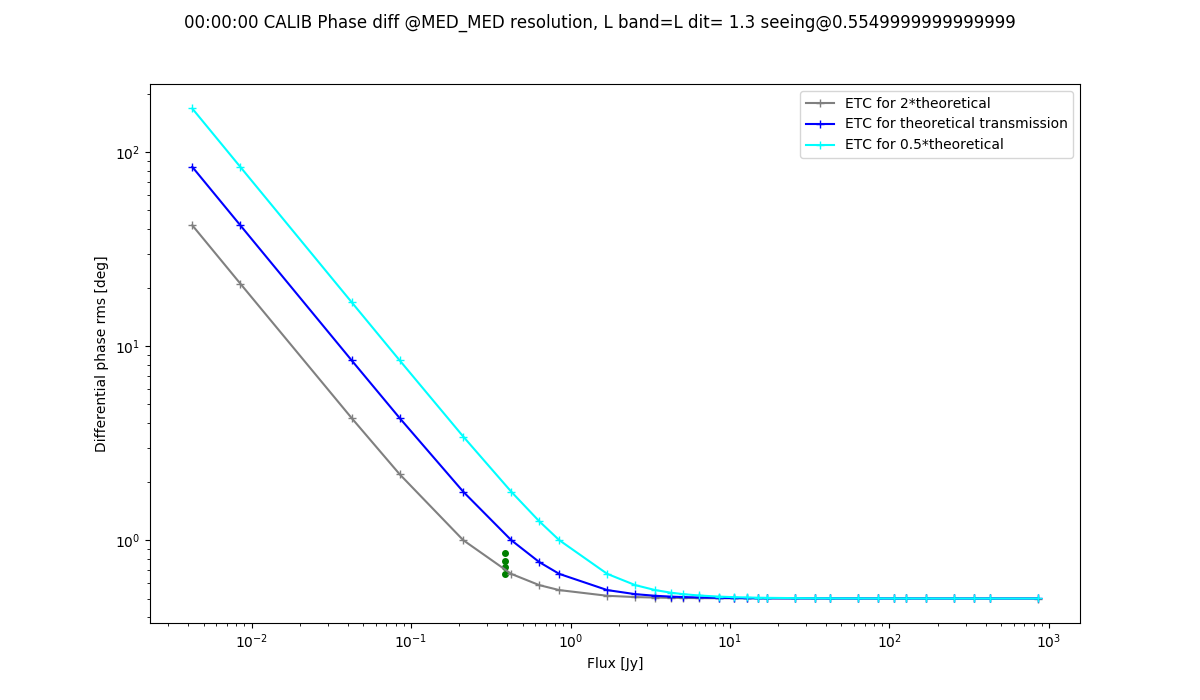}\\
  \includegraphics[width=0.5\textwidth]{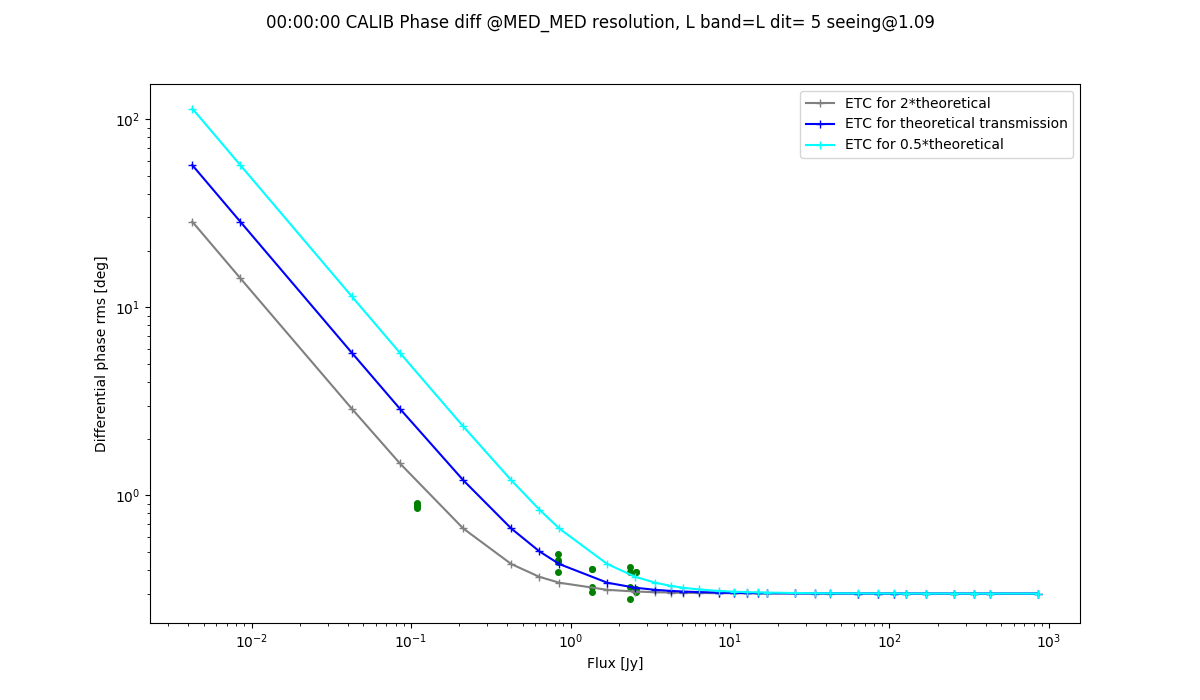}\hfill
  \includegraphics[width=0.5\textwidth]{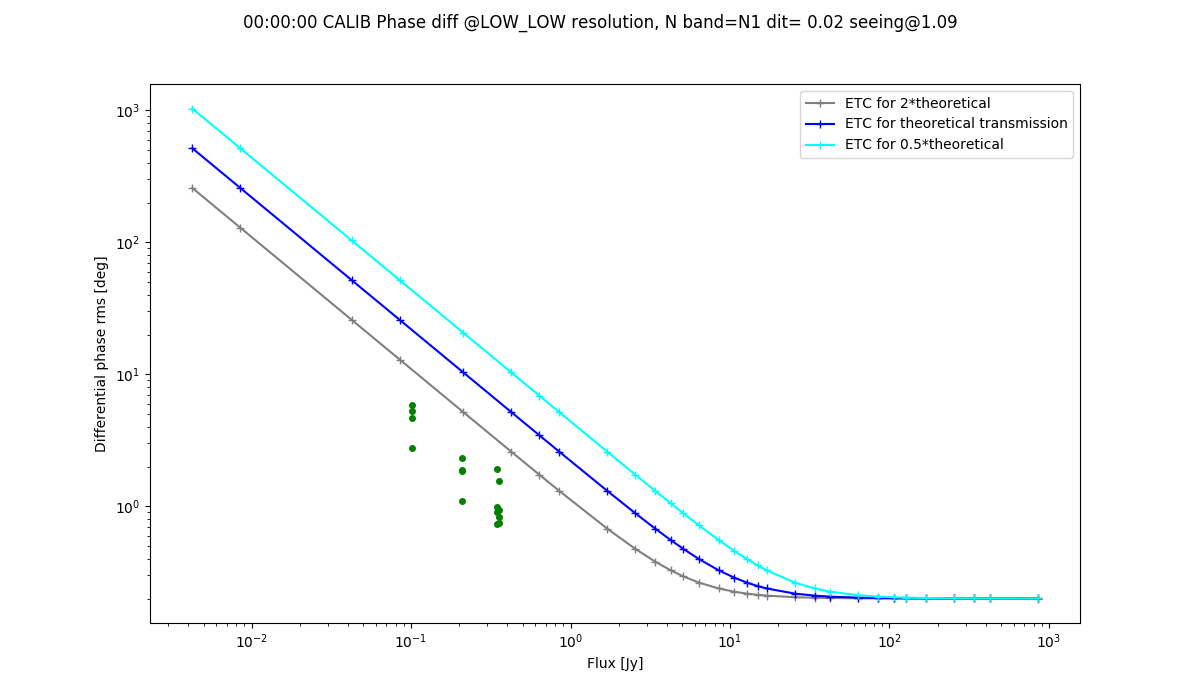}
  \caption{Differential phase RMS per spectral channel for a one-minute
  observation, as a function of the flux of the source. Each point is a
  one-minute exposure; the curves are ETC predictions for half, one and twice
  the theoretical transmission. \emph{Top left:} LOW in LM, DIT $=1.3$\,s.
  \emph{Top right:} MED in LM, DIT $=1.3$\,s. \emph{Bottom left:} MED in LM,
  December run, DIT $=5$\,s. \emph{Bottom right:} LOW in N, 8.2--8.9\,$\mu$m,
  DIT $=0.02$\,s.}
  \label{fig:senslm}
\end{figure}

In LM LOW resolution (Fig.~\ref{fig:senslm}, top left) two faint calibrators with
nearby tip-tilt stars follow the ETC predictions, which confirms the offered
L-band correlated flux limit with GPAO-LGS.
In LM MED (Fig.~\ref{fig:senslm}, top right) the single November calibrator
HD~285851 ($L \simeq 400$\,mJy) is brighter than the 40\,mJy correlated flux
limit of that mode, and also follows the ETC.

We extended the test in the December run to fainter sources and longer
integrations (Fig.~\ref{fig:senslm}, bottom). Several faint N-band calibrators observed in MED LM
with DIT $=5$\,s still follow the ETC, so the MED LM performances are maintained
with the laser guide stars. Conditions were poor that night, with
$\tau_0 < 2$\,ms, which degraded the GRAVITY fringe tracker and produced many
fringe jumps; most were recovered quickly, and processing the LM data without
fringe-jump detection gave a better signal-to-noise ratio and more stable
visibilities. Those are the data shown here.

In N-band LOW resolution, several calibrators between 0.1 and 0.4\,Jy come out
better than the ETC predictions. The differential phase error per spectral
channel and per minute on the 0.1\,Jy source lies between 2 and 4$^\circ$, and
4$^\circ$ corresponds to a signal-to-noise ratio of 10 on the correlated flux,
so the offered N-band correlated flux limit is reached with the LGS. Because
the GRAVITY fringe tracker data were too unstable that night to be used in the
coherent processing, the N-band frames were cophased on the N-band data
themselves.

\subsection{Fringe-tracker stability}
\label{sec:lgsft}

\begin{figure}[htbp]
  \centering
  \includegraphics[width=0.9\textwidth]{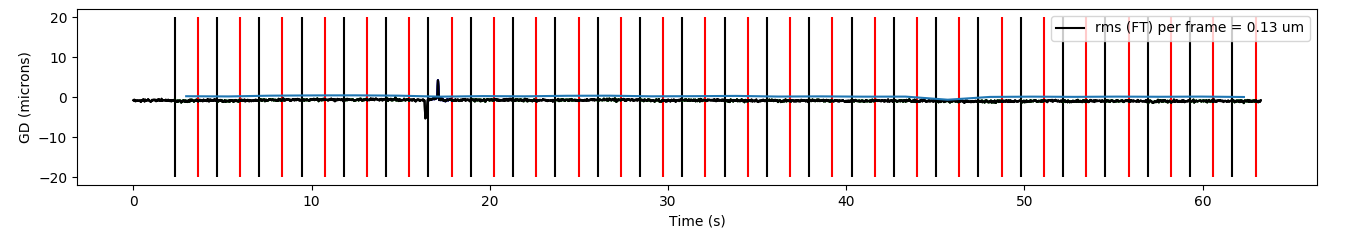}\\
  \includegraphics[width=0.9\textwidth]{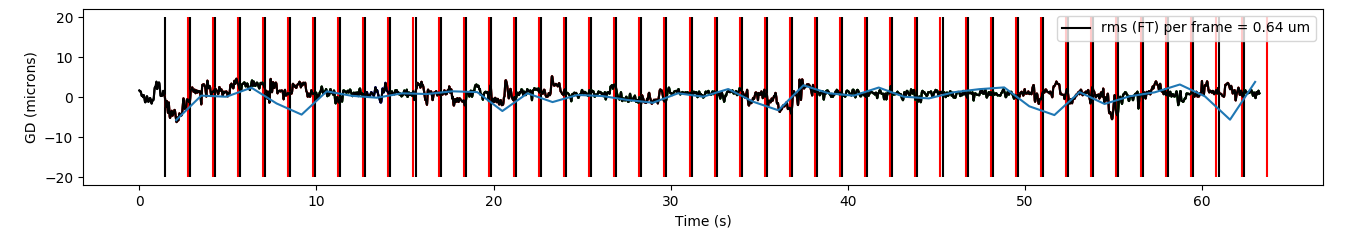}
  \caption{GRAVITY fringe tracker OPD (black) and the a posteriori estimated
  MATISSE OPD (blue); the vertical black and red lines mark the beginning and
  end of a MATISSE frame. The mean FT residual per frame is given for each
  baseline. \emph{Top (bright case):} HD~285851, $\sim100$\,nm RMS. \emph{Bottom (faint case):}
  HD~294236, $\sim600$\,nm RMS.}
  \label{fig:lgsft}
\end{figure}

We extracted the GRAVITY fringe tracker residuals over the duration of each
MATISSE frame (Fig.~\ref{fig:lgsft}). On the bright cophasing calibrator
HD~285851 ($K=7.2$, $L=0.4$\,Jy) the residual is about 100\,nm RMS per frame,
with a single quick fringe jump over the exposure. On HD~294236 ($K=11.3$,
$L=0.01$\,Jy), close to the fringe-tracking limit, it rises to about 600\,nm
RMS. The bright-star value is consistent with the 180\,nm RMS accuracy
originally specified for an external fringe tracker; the faint-star value quantifies the
degradation to be expected when observing at the limit.

\subsection{Transfer function stability}
\label{sec:lgstf}

\begin{figure}[htbp]
  \centering
  \includegraphics[width=0.85\textwidth]{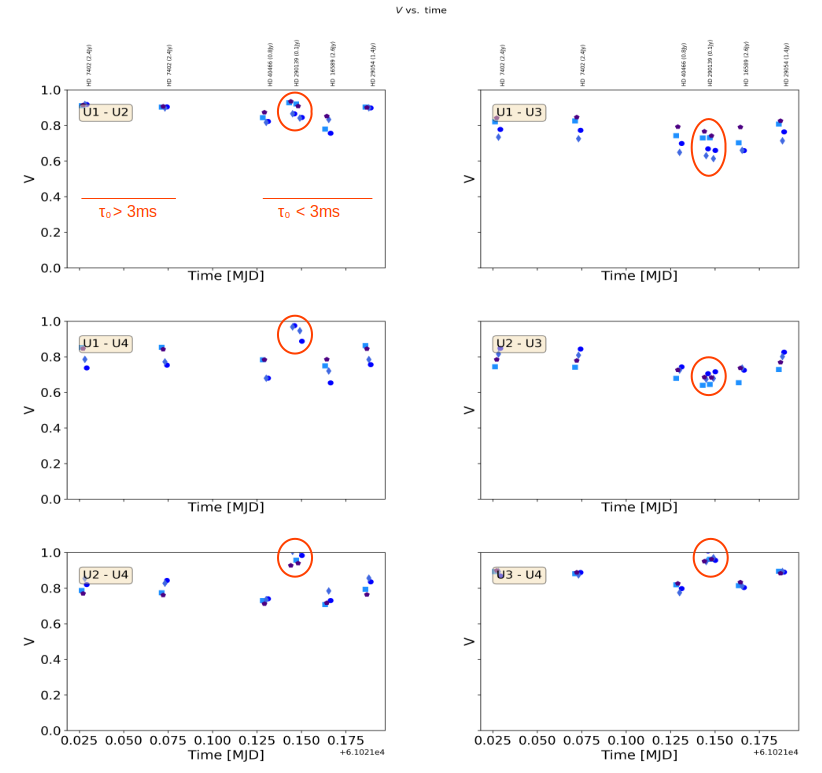}
  \caption{L-band transfer function from MED-resolution observations without
  chopping, night of 11--12 December. Each point is the instrumental visibility
  averaged over 3.5--3.85\,$\mu$m for a one-minute exposure; each symbol is a
  different BCD configuration. The circled point is the faintest target
  ($L=0.1$\,Jy) and is excluded from the stability analysis.}
  \label{fig:lgstf}
\end{figure}

We assessed the stability of the L-band transfer function on the night of
11--12 December, using the calibrators originally selected for the N-band
sensitivity tests (Fig.~\ref{fig:lgstf}). These were observed without chopping,
so residuals from the background subtraction degraded the accuracy of the
absolute visibilities on the faintest targets, below about 1\,Jy.

For a given BCD position and fair conditions ($\tau_0 > 3$\,ms), the transfer
function is stable at the level of a few percent between observing blocks
separated by more than an hour. In poor conditions ($\tau_0 < 3$\,ms) it
dropped for two calibrators because of the degraded fringe tracking, while
remaining stable between them at about 5\% for a given BCD configuration. The
last calibrator of the night, HD~7402, agrees closely with the first two
observations despite the still poor $\tau_0$.

\subsection{Strehl ratio: IRIS against MATISSE}
\label{sec:lgsstrehl}

\begin{figure}[htbp]
  \centering
  \includegraphics[width=0.8\textwidth]{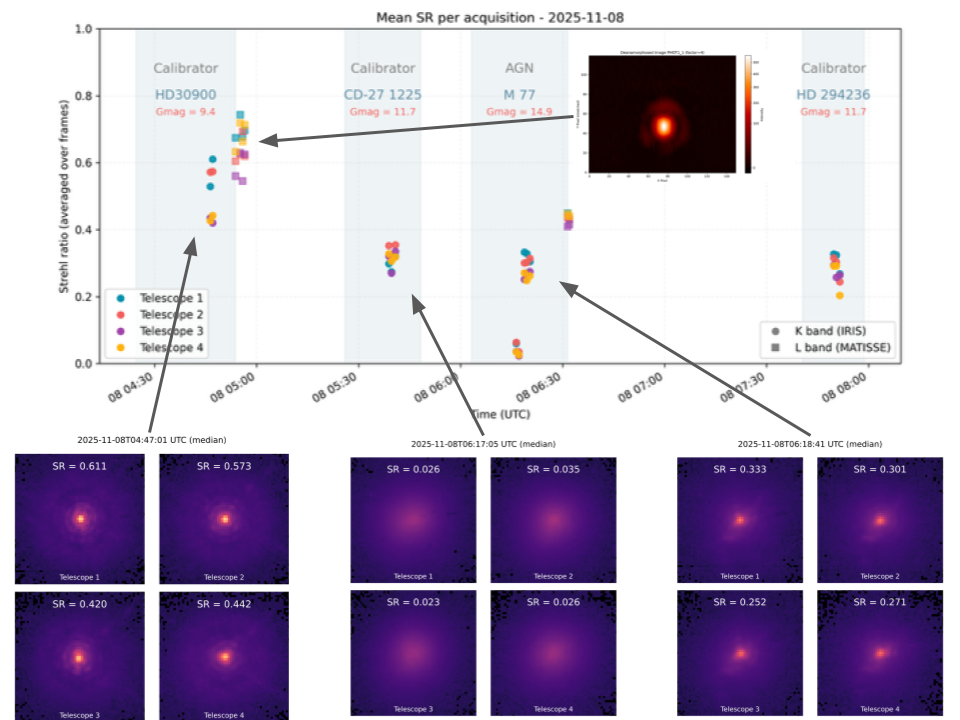}
  \caption{Strehl ratio measured on IRIS acquisition images (round symbols, K
  band) and on MATISSE acquisition images (square symbols, L band), per
  telescope. Shaded regions mark the time intervals of the observed targets,
  labelled with their G-band magnitude.}
  \label{fig:lgsstrehl}
\end{figure}

On the night of 8 November 2025 we obtained acquisition images with both IRIS
and MATISSE, which allows us to compare the Strehl ratio delivered by GPAO in
the two bands (Fig.~\ref{fig:lgsstrehl}). The mean Strehl per acquisition,
averaged over all frames and separated by telescope, gives a ratio of 1.4
between the L-band and the K-band Strehl.

\subsection{Calibrator strategy with the laser guide stars}
\label{sec:lgscal}

One difficulty arises for MATISSE with the LGS. Even for the faintest
N-band science targets, the N-band calibrators have $G_{\rm RP} < 8.5$, that is
they are brighter than the lasers. When such a calibrator is used as the
tip-tilt star, which is the usual practice, and the laser is pointed close to
it, the adaptive optics is confused. One remedy is to point the laser away from
the calibrator, by more than 15$''$ for $6.5 < G_{\rm RP} < 8.5$ and by more
than 20$''$ for $5.0 < G_{\rm RP} < 6.5$; below $G_{\rm RP} = 5.0$ the
wavefront sensor saturates. This strategy was used successfully on the night of 11--12 December, where the
night log records the laser offset by 15$''$ for HD~7402 and HD~40466 and by
20$''$ for HD~16589 and HD~29054, after an over-illumination alarm on UT4
attributed to the $G_{\rm RP} = 6.8$ central star of HD~7402 at a 10$''$
separation.

The transfer functions of HD~7402 in Fig.~\ref{fig:lgstf}, however, show no
observable difference between the first observing block, taken with NGS, and
the second, taken with LGS and the laser pointed 15$''$ away. We therefore
recommend that all calibrators brighter than $G_{\rm RP} = 8.5$ be observed
with the natural guide star mode, even when the laser guide stars are used on
the science target.

\subsection{MATISSE with laser guide stars: estimated performances}

With GPAO in LGS-VIS mode (Table~\ref{tab:lgs}), the L, M and N band limits are
confirmed at the values already reached in NGS. The exception is the L-band
HIGH+ mode, whose limit improves by a factor of 6, from 6\,Jy to 1\,Jy, a gain
that also applies to the GPAO NGS mode. What the LGS mode adds lies elsewhere. It removes the constraint on the visible
brightness of the target itself, so that these limits become reachable on red
and embedded sources for which no suitable natural guide star exists.

\begin{table}[htbp]
  \centering
  \caption{MATISSE Period~117 sensitivity limits, GRA4MAT with the UTs, GPAO
  LGS-VIS. Fluxes in Jy.}
  \label{tab:lgs}
  \resizebox{\textwidth}{!}{%
\begin{tabular}{lccccccc}
    \toprule
    \multirow{2}{*}{Spectral mode} & \multicolumn{2}{c}{Lowest L flux (Jy)}
      & \multicolumn{2}{c}{Lowest M flux (Jy)}
      & \multicolumn{2}{c}{Lowest N flux (Jy)} \\
    \cmidrule(lr){2-3}\cmidrule(lr){4-5}\cmidrule(lr){6-7}
      & Corr. flux \& diff. phase & Closure phase
      & Corr. flux \& diff. phase & Closure phase
      & Corr. flux \& diff. phase & Closure phase \\
    \midrule
    LOW ($R\sim35$)    & 0.01 (confirmed) & 0.03 & 0.03 & 0.04
                       & 0.1 (confirmed)  & 0.1 \\
    MED ($R\sim500$)   & 0.04 (confirmed) & 0.06 & 0.35 & 0.45
                       & not offered      & not offered \\
    HIGH ($R\sim1000$) & 0.12             & 0.16 & not offered & not offered
                       & 0.5 ($R\sim220$) & 0.7 ($R\sim220$) \\
    HIGH+ ($R\sim3300$)& $6 \rightarrow 1$ ($\times$6) & 8 & 6 & 8
                       & not offered      & not offered \\
    \bottomrule
  \end{tabular}}
\end{table}

\subsection{Work in progress}

Solar system targets and the narrow off-axis mode were not attempted during
these two runs, and the imaging capability and the standalone mode remain to be
verified.

\section{MATISSE SCIENCE IN THE GRAVITY+ ERA}
\label{sec:science}

The gain in performances reported above applies to an instrument that is already
scientifically productive on a wide range of topics\cite{Petrov2024}. Recent results include the
imaging of the disk of Z~CMa during outburst\cite{Lykou2025}, the
characterisation of the gas envelopes of Cepheids\cite{Hocde2025}, the
localisation of the black hole in NGC~1068\cite{GamezRosas2022, Leftley2024},
the detection of iron-rich dust in the rocky-planet-forming region of the
multi-ringed disk of HD~144432\cite{Varga2024}, the dust engine of the Circinus
galaxy\cite{Isbell2022, Isbell2023}, the imaging of the environment of the
companion of $\pi^1$~Gruis\cite{Drevon2026}, the dusty spiral arcs around
3~Puppis\cite{Abello2025}, and the mid-infrared spectrum of the exoplanet
$\beta$~Pic~b\cite{Houlle2025}.

What MATISSE contributes here is a wide instantaneous bandwidth, which allows
the mapping of mineralogy and molecular features, and a position between the
millimetre domain of ALMA and the shorter wavelengths of PIONIER and GRAVITY,
already at the angular resolution that ELT/METIS will explore. GPAO makes it
possible to image AGN cores at parsec scale, to follow and image transient
events, and to monitor dust processing in real time. The most promising
follow-ups are low-mass young stellar objects, exoplanets and AGN, which are
also of interest for the cosmological distance ladder and the Hubble tension. We show on a few examples the new capabilities of MATISSE with the GPAO AO.

\subsection{A few illustrations of the new MATISSE capabilities}

\subsubsection{Faint-target pipeline outputs}
\label{sec:lgsfaint}

\begin{figure}[htbp]
  \centering
  \includegraphics[width=0.8\textwidth]{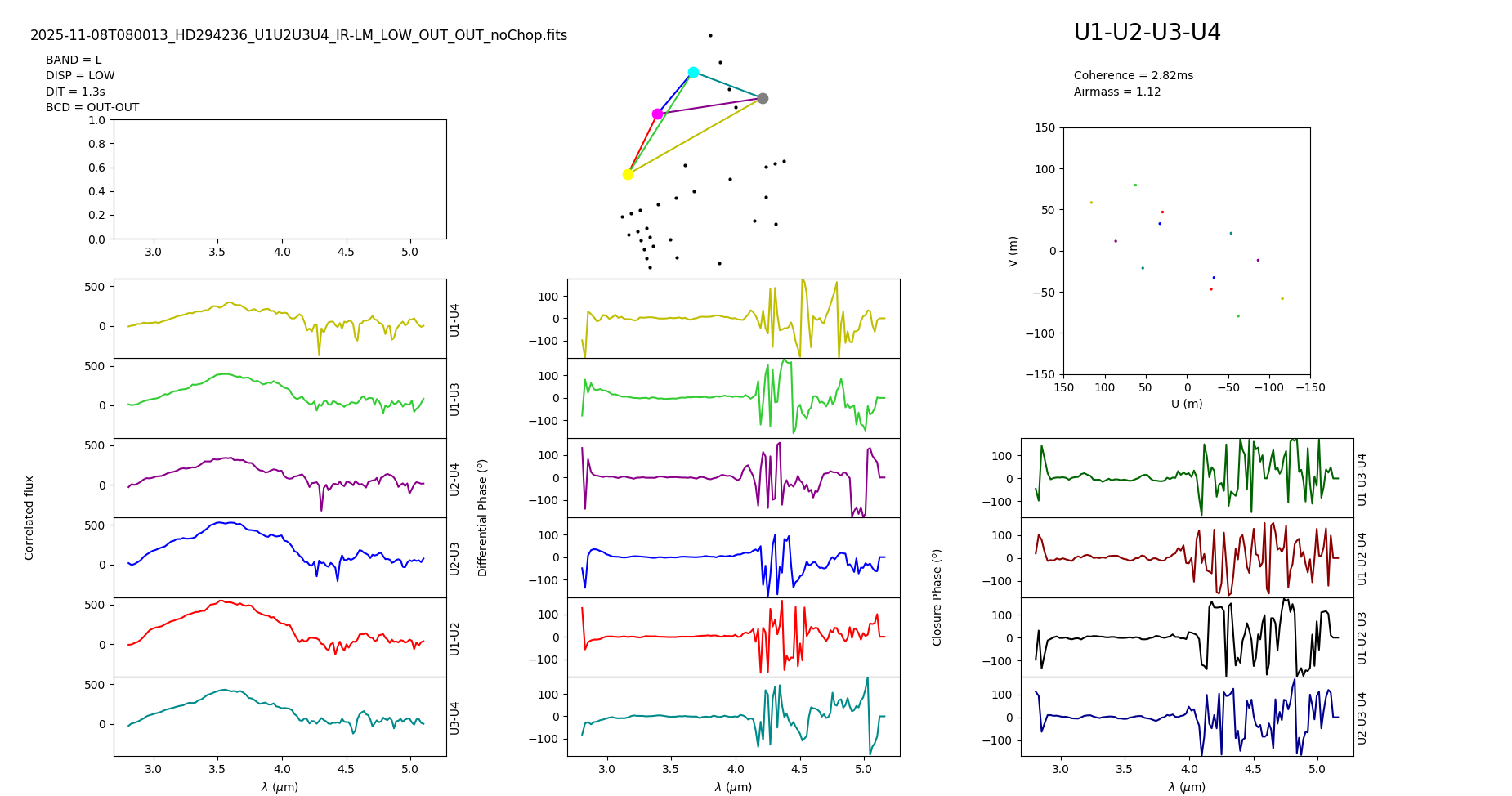}\\
  \includegraphics[width=0.8\textwidth]{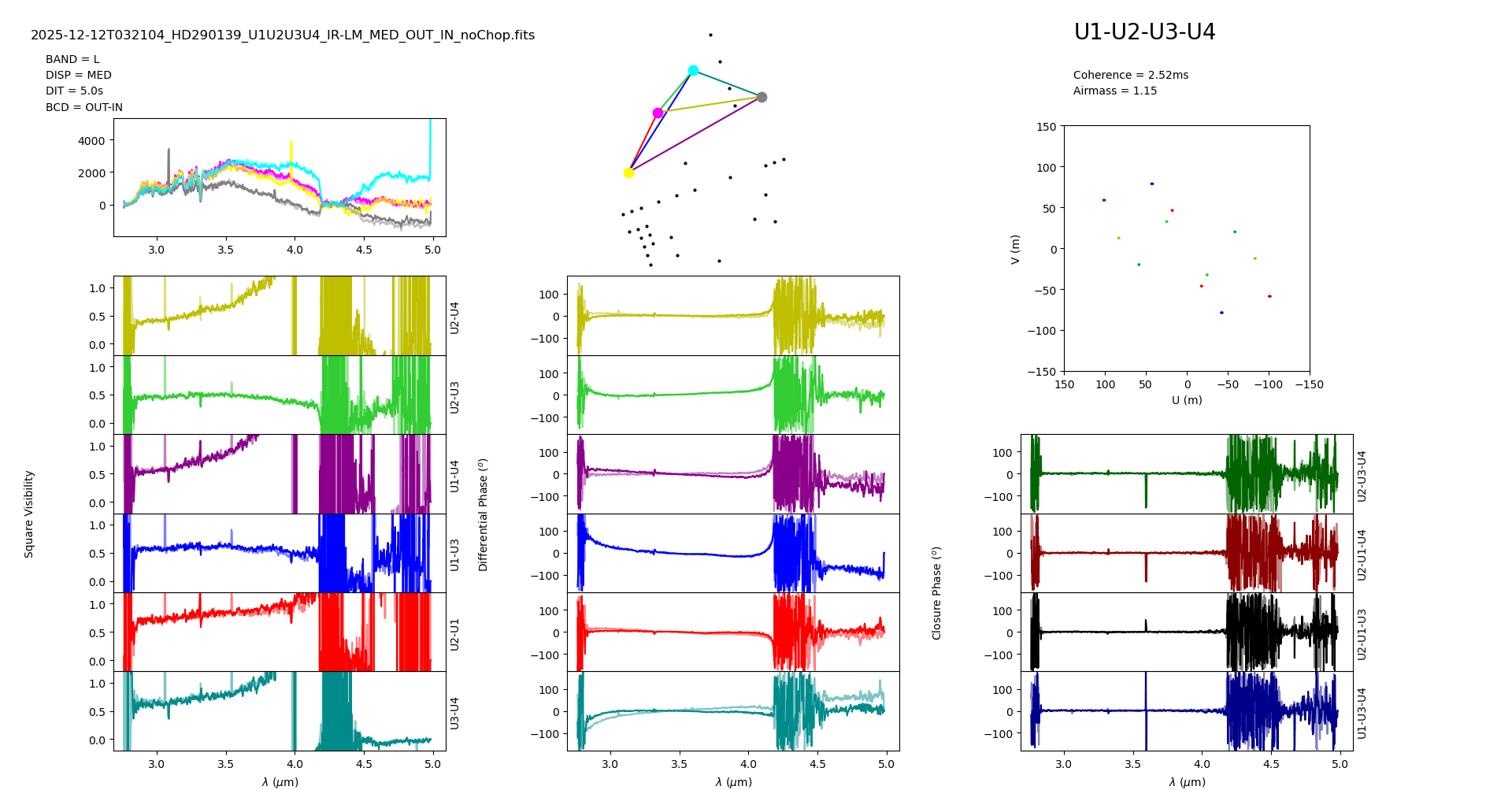}\\
  \includegraphics[width=0.8\textwidth]{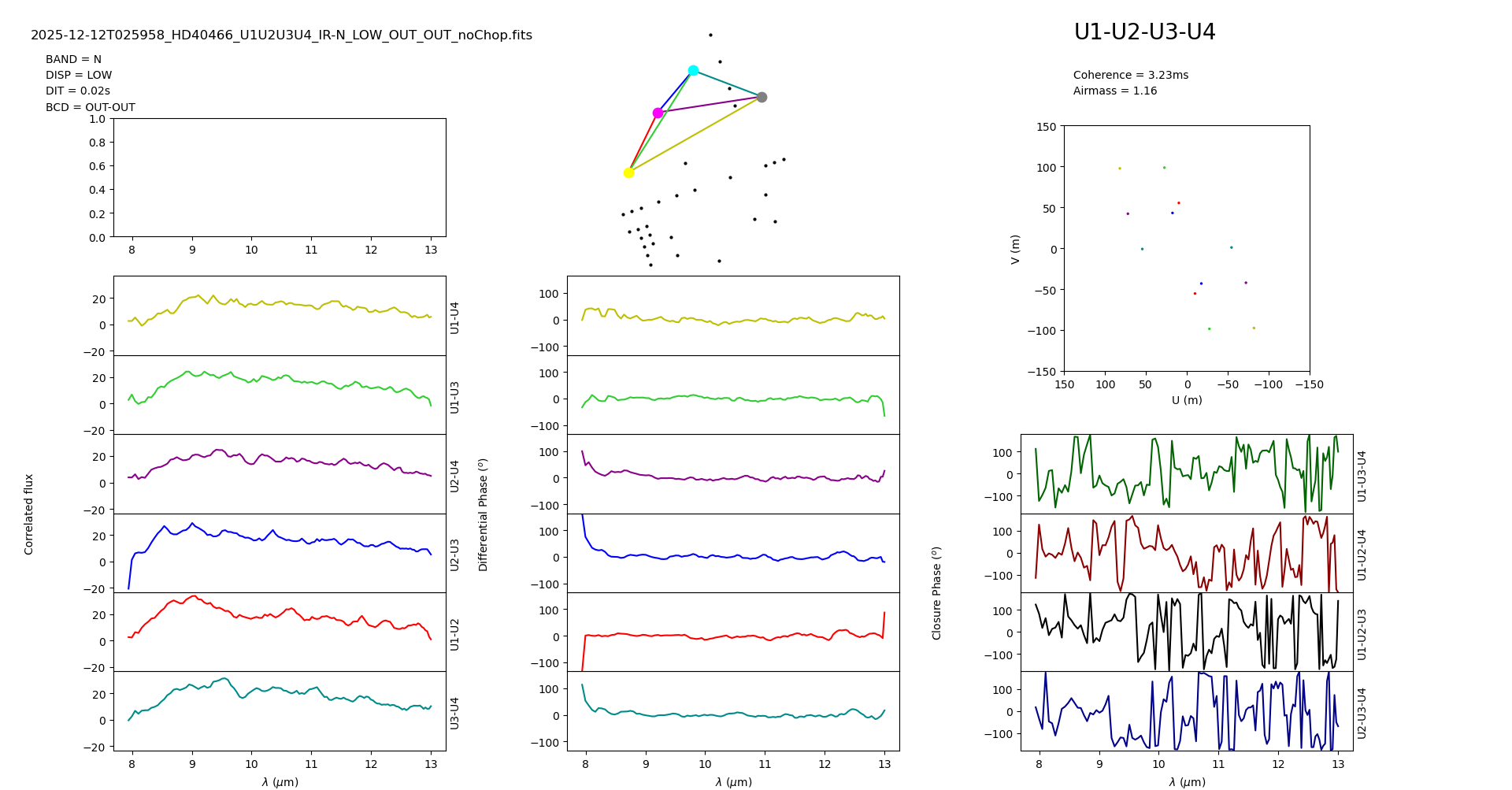}
  \caption{Typical MATISSE/GRA4MAT pipeline outputs. \emph{Top:} HD~294236, 10\,mJy, LM LOW.
  \emph{Middle:} HD~290139, 100\,mJy, LM MED. \emph{Bottom:} HD~40466,
  0.1\,Jy, N LOW.}
  \label{fig:lgsdrs}
\end{figure}

Figure~\ref{fig:lgsdrs} shows the data reduction software outputs at three
points along the flux range covered by the campaign: HD~294236 at 10\,mJy in
LM LOW, the faintest target of the campaign; HD~290139 at 100\,mJy in LM MED;
and HD~40466 at 0.1\,Jy in N LOW. Correlated fluxes, differential phases and
closure phases are recovered in all three.

\subsubsection{Young stellar objects}
\label{sec:lgsyso}

We observed two embedded young stellar objects: V806~Tau (Haro~6-13) in MED,
and HL~Tau ($G_{\rm RP} = 13$, $L = 2.5$\,Jy) in MED and VHIGH (HIGH+), the first
MATISSE observations of that target in either mode. Neither could be observed
before GPAO, because the visible wavefront-sensing magnitude limited them well
before their mid-infrared brightness did.

\begin{figure}[htbp]
  \centering
  \includegraphics[width=0.48\textwidth]{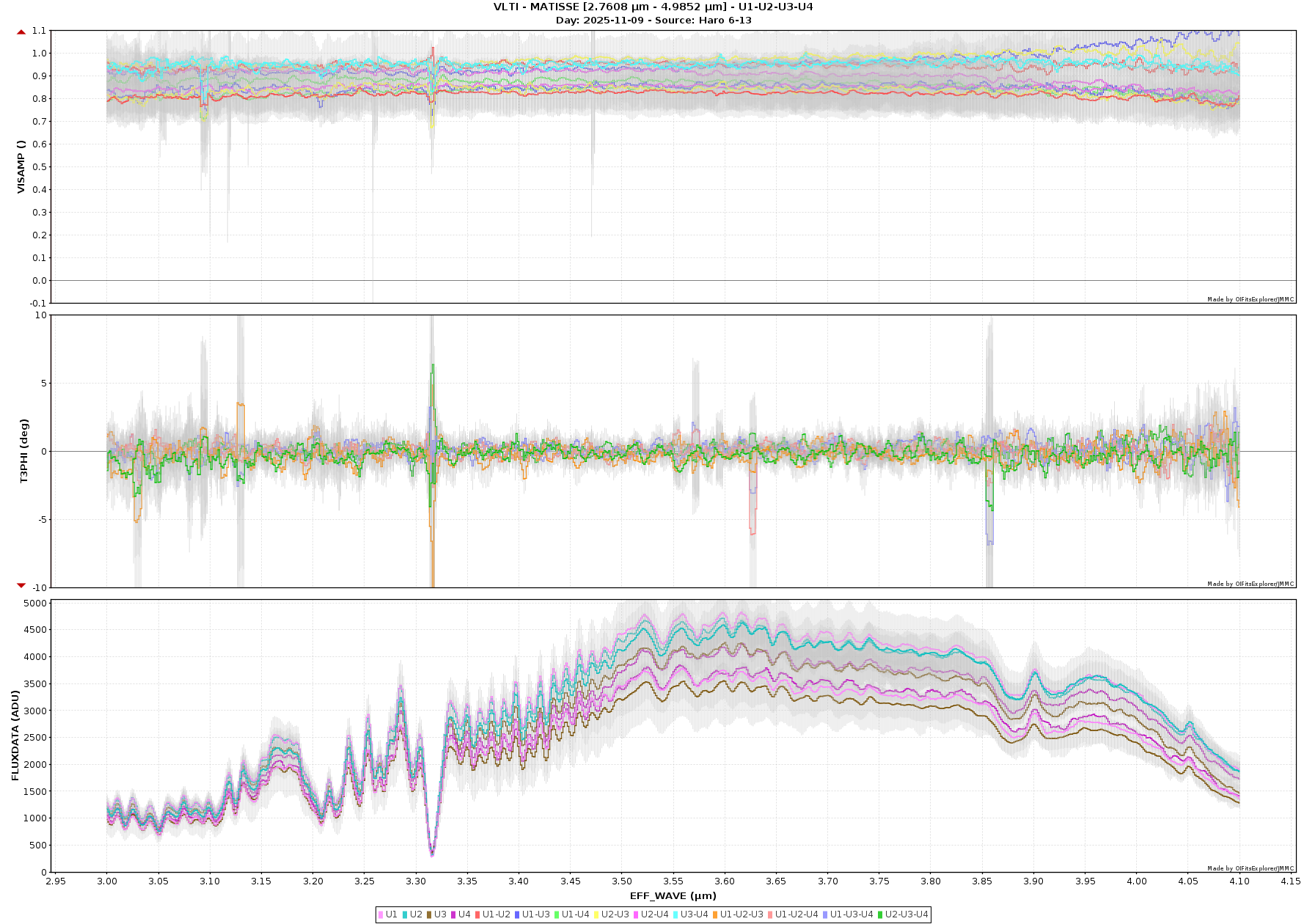}\hfill
  \includegraphics[width=0.48\textwidth]{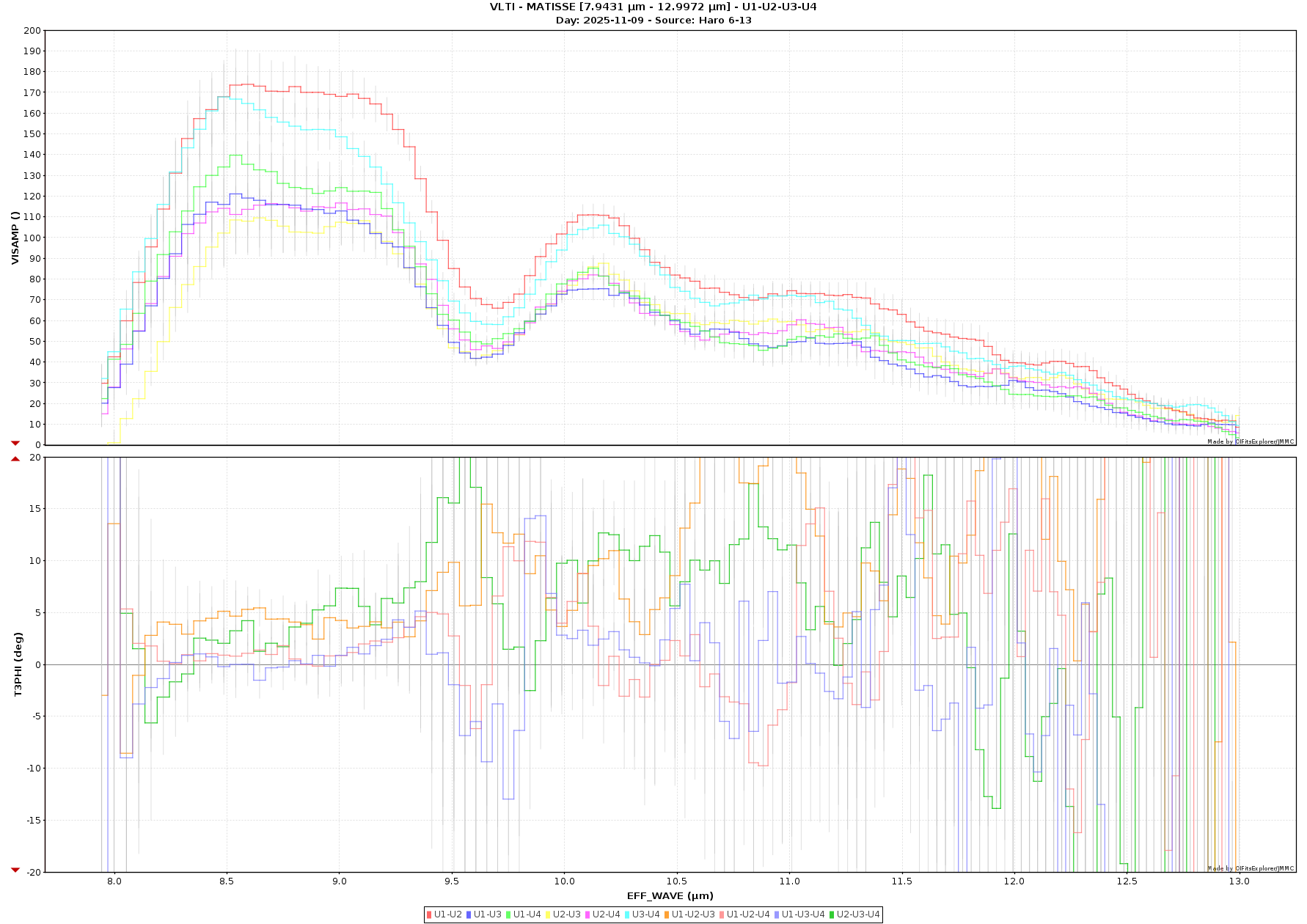}\\
  \includegraphics[width=0.48\textwidth]{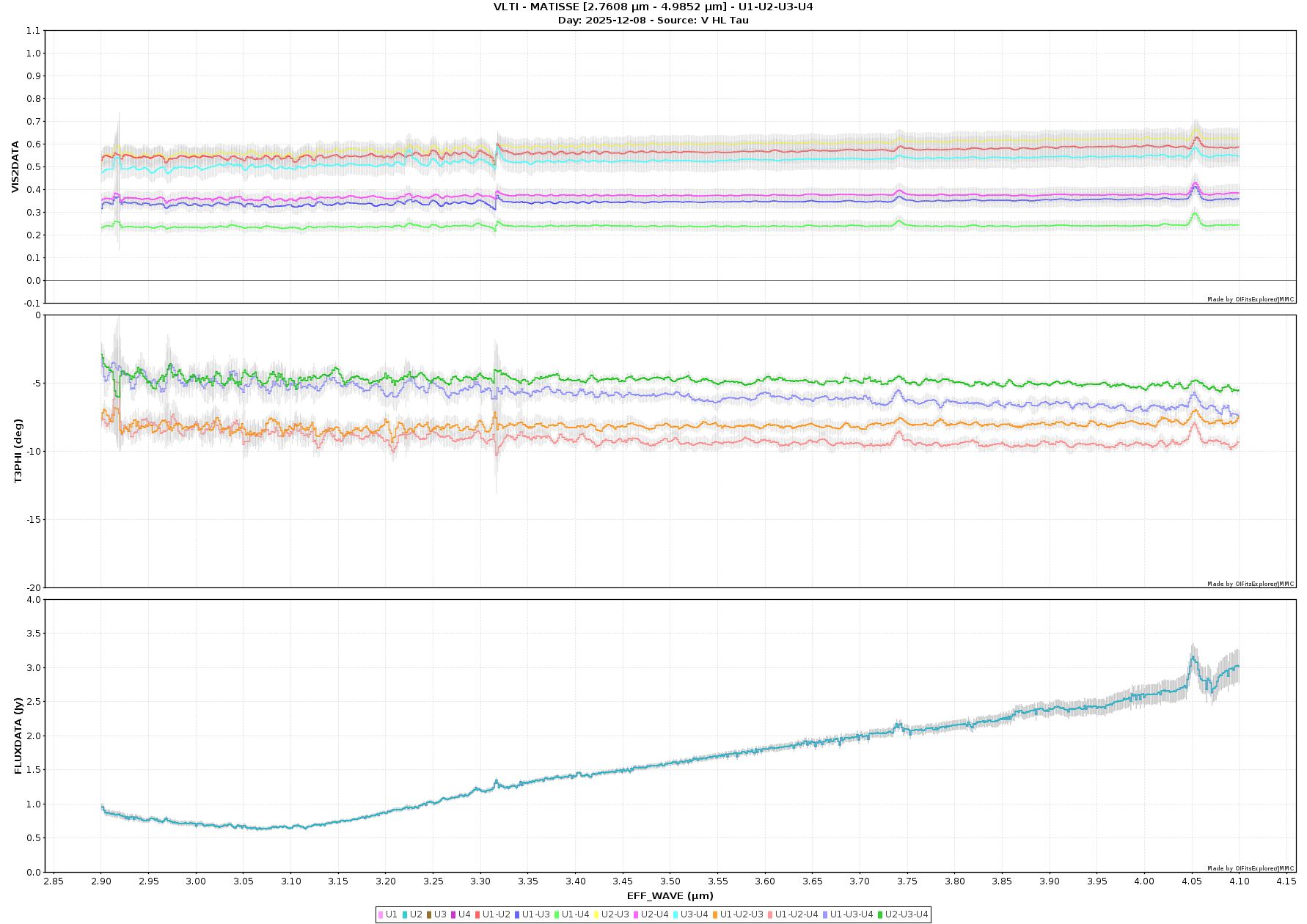}\hfill
  \includegraphics[width=0.48\textwidth]{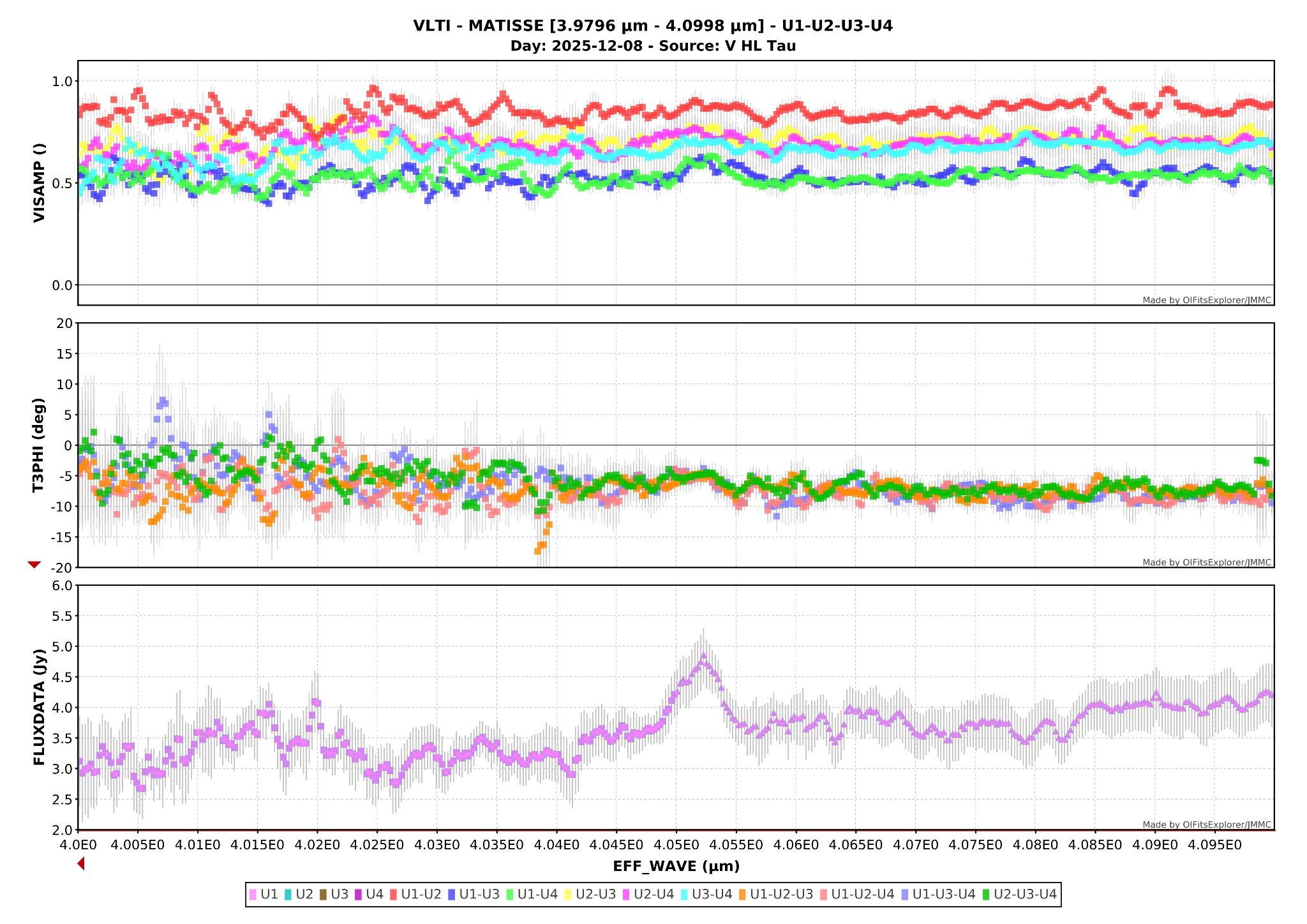}
  \caption{Young stellar objects. \emph{Top left:} V806~Tau, calibrated LM-band
  visibilities (chopped), calibrated closure phases (non-chopped) and
  uncalibrated total spectrum (chopped) in MED, BCD-merged. \emph{Top right:}
  V806~Tau, uncalibrated correlated fluxes and closure phases in the N band.
  \emph{Bottom left:} HL~Tau in MED. \emph{Bottom right:} HL~Tau in HIGH+,
  both non-chopped and calibrated, with the chopped total spectrum.}
  \label{fig:v806tau}
\end{figure}

The LM-band data quality on V806~Tau was good, and visibilities, phases and
closure phases could be computed. The Br$\alpha$ line is seen in emission in
the total spectrum (Fig.~\ref{fig:v806tau}, top left). In the N band the calibrator
of V806~Tau was too faint ($N = 0.04$\,Jy) for a reliable absolute visibility
and closure phase calibration, so Fig.~\ref{fig:v806tau} (top right) shows the
uncalibrated correlated fluxes and closure phases. Despite the faintness of the
target in N ($N = 0.5$\,Jy), the correlated fluxes have the shape typical of
uncalibrated N-band MATISSE spectra and the closure phases are usable between 8
and 9\,$\mu$m.

On HL~Tau the LM-band data quality was good in both MED and HIGH+
(Fig.~\ref{fig:v806tau}, bottom). Only the non-chopped fringe data proved usable in LM:
chopping gave poor fringe measurements because of the adverse conditions
($\tau_0 < 3$\,ms) during the observation, and in HIGH+ only the IN-IN
exposure was usable. The hydrogen lines of interest are detected at both
resolutions, and the HIGH+ total spectrum shows a line that starts to be
spectrally resolved.

\subsubsection{Embedded AGN in medium spectral resolution: NGC 1068}
\label{sec:lgsagn}

\begin{figure}[htbp]
  \centering
  \includegraphics[width=0.8\textwidth]{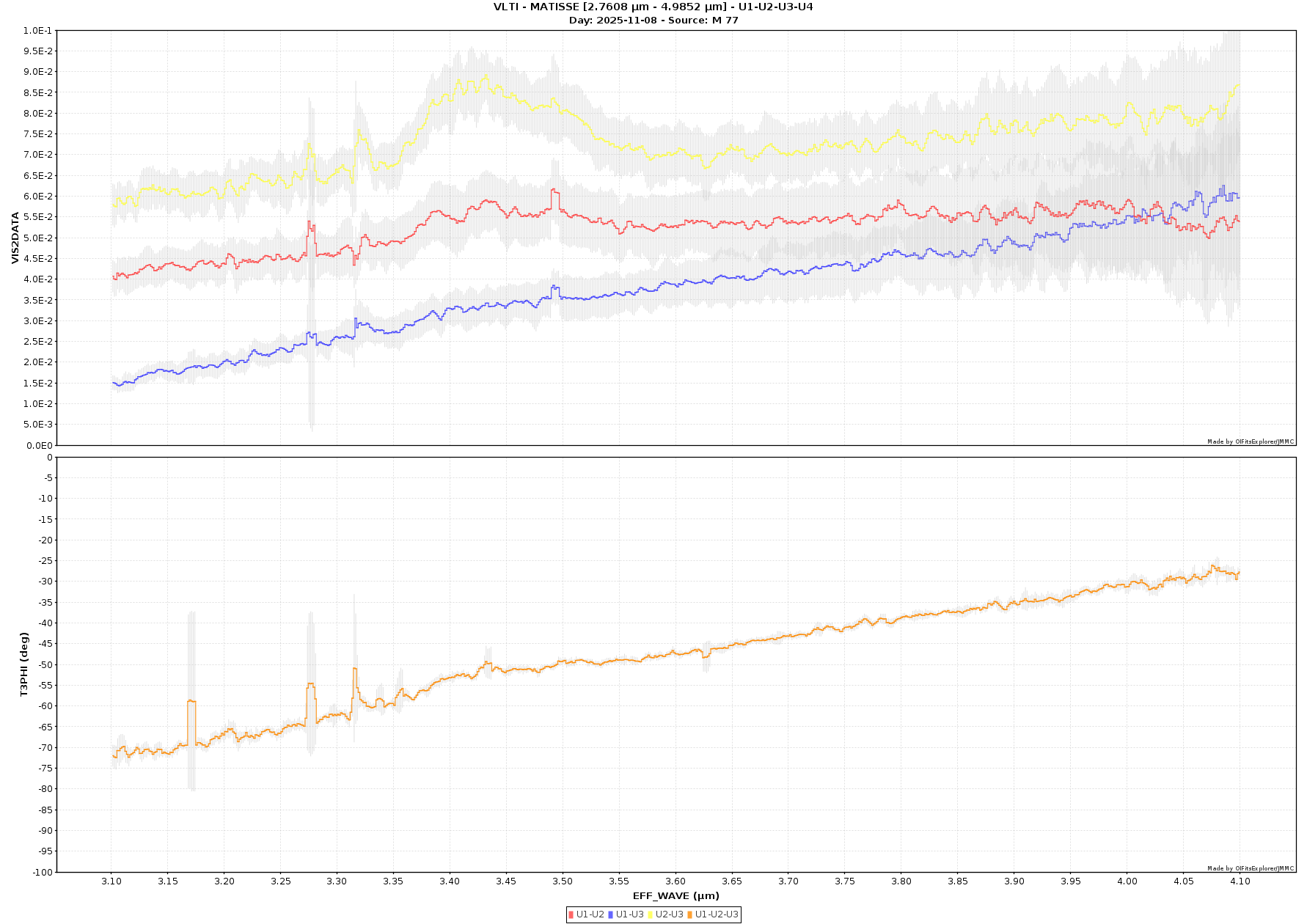}
  \caption{Visibilities and closure phase of NGC~1068 on the three baselines
  where fringe tracking was possible, at medium spectral resolution.}
  \label{fig:lgsngc}
\end{figure}

Fringe locking on NGC~1068 succeeded on three baselines; GRA4MAT failed to find
fringes on the other three. Data were recorded on the three usable baselines
and were sufficient to compute visibilities and one closure phase
(Fig.~\ref{fig:lgsngc}). They show the carbonaceous signature previously
detected in the spectrum of this source\cite{GamezRosas2022}.
This is the first NGC~1068 dataset at medium spectral resolution, and the first
on this target with GRA4MAT; it demonstrates that this resolution is feasible on
faint targets of $G_{\rm RP} \simeq 11.5$.

\section{CONCLUSIONS}
\label{sec:conclusions}

MATISSE with GPAO keeps the same instrumental transfer function level, gains
in transfer-function stability in the L band, gains 20--30\% in correlated flux
in GRA4MAT mode, and receives a factor 1.3--1.4 more flux in the L band at low
resolution, and nothing in the N band. In terms of offered limits, the
improvement is at least a factor of two in sensitivity in the L and M bands, up
to a factor of six or seven in the dispersed modes, and one magnitude on the
fringe-tracking limit. The laser guide stars also decouple the wavefront
sensing from the visible brightness of the science target, which brings red and
embedded objects such as HL~Tau and NGC~1068 within reach for the first time.
The 2025 recommissioning confirms that the instrument behaves as predicted
with the laser guide stars. The offered correlated flux limits are reached in
LM LOW and MED and exceeded in N LOW, the transfer function is stable at the
few percent level, and the L-band Strehl ratio delivered by GPAO is 1.4 times
the K-band one. On the operational side, calibrators brighter than
$G_{\rm RP}=8.5$ are best observed with the natural guide stars even when the
science target uses the lasers. Solar system targets,
the narrow off-axis mode, the imaging capability and the standalone mode remain
to be verified.

The mid-infrared arm of the VLTI therefore still has margin for improvement, which supports the study of an upgraded instrument, MATISSE+, designed to exploit the wavefront quality that GPAO now delivers routinely.

\acknowledgments

MATISSE has been built by a consortium led by the Observatoire de la C\^ote
d'Azur (France) together with Max-Planck institutes (MPIfR, MPIA, Germany),
Leiden Observatory and SRON (Netherlands), and contributions from Konkoly
observatory, Kiel university and Cologne university. GRAVITY+ is a consortium
composed of German (MPE, MPIA, University of Cologne), French (CNRS-INSU:
LESIA Paris, IPAG Grenoble, Lagrange Nice, CRAL Lyon), British (University of
Southampton), Belgian (KU Leuven) and Portuguese (CAUP) institutes, built in
close collaboration with ESO.

This work was supported by the \textit{Agence Nationale de la Recherche} (ANR)
through the EXOVLTI (ANR-21-CE31-0017), MASSIF (ANR-21-CE31-0018) and AGN
MELBA (ANR-21-CE31-0011) projects. French partners thank the specific action
ASHRA and the national programs PNP, PNPS and PNCG.

The results of Sect.~\ref{sec:lgs} rest on nights shared with the ESO and
GRAVITY+ commissioning teams at Paranal and in Garching, whom we thank for the
organization, the observations and the night logs.

This work is based on observations collected at the European Southern
Observatory under  during commissionning of the GRAVITY+ instrument, and espacially its adaptive optics GPAO.
This research has made use of the SIMBAD database and of the VizieR catalogue
access tool, operated at CDS, Strasbourg, France, and of the Paranal
Astronomical Site Monitor archive. JMMC tools were also used for this work,
namely ASPRO2, searchCal, searchFTT and oifitsexplorer.


\end{document}